# Automatic reconstruction of fault networks from seismicity catalogs including location uncertainty

Y. Wang,[1] G. Ouillon,[2] J. Woessner,[1] D. Sornette,[3] S. Husen[1]

[1] Swiss Seismological Service, ETH Zurich, Zurich, Switzerland

[2] Lithophyse, 4 rue de l'Ancien Sénat, 06300 Nice, France

[3] Department of Management, Technology and Economics, ETH Zurich, Zurich, Switzerland

**Abstract**: We introduce the Anisotropic Clustering of Location Uncertainty Distributions (ACLUD) method to reconstruct active fault networks on the basis of both earthquake locations and their estimated individual uncertainties. After a massive search through the large solution space of possible reconstructed fault networks, we apply six different validation procedures in order to select the corresponding best fault network. Two of the validation steps (cross-validation and Bayesian Information Criterion (BIC)) process the fit residuals, while the four others look for solutions that provide the best agreement with independently observed focal mechanisms. Tests on synthetic catalogs allow us to qualify the performance of the fitting method and of the various validation procedures. The ACLUD method is able to provide solutions that are close to the expected ones, especially for the BIC and focal mechanism-based techniques. The clustering method complemented by the validation step based on focal mechanisms provides good solutions even in the presence of a significant spatial background seismicity rate. Our new fault reconstruction method is then applied to the Landers area in Southern California and compared with previous clustering methods. The results stress the importance of taking into account undersampled sub-fault structures as well as of the spatially inhomogeneous location uncertainties.

## 1 Introduction

Earthquake forecasts should ultimately be founded on the premise that seismicity and faulting are intimately interwoven: earthquakes occur on faults and faults grow and organize in complex networks through accumulation of earthquakes. The obvious character and the power of this well-established fact are obfuscated by serious difficulties in exploiting it for a better science of earthquakes and their prediction. Indeed, an intrinsic limitation of present efforts to forecast earthquakes lies in the fact that only a limited part of the full fault network has been revealed, notwithstanding the best efforts combining geological, geodetic and geophysical methods (see *Mace and Keranen* [2012], for instance) together with past seismicity to illuminate fault structures [*Plesch et al.*, 2007]. Nevertheless, these studies suggest that fault networks display multiscaling hierarchical properties [*Cowie et al.*, 1995], which are intimately associated with the modes of tensorial deformations accommodating large scale tectonic driving forces [*Sornette*, 1991; *Sornette and Virieux*, 1992]. Neglecting the information from fault networks constitutes a major gap in the understanding of the spatial-temporal organization of earthquakes (see however early attempts by *Cowie et al.* [1995]; *Cowie et al.* [1993]; and *Sornette et al.* [1994]), thus limiting the quality and efficiency of most current earthquake forecasting methods. Including more realistic geometries and tensorial strain information associated with the underlying reconstructed fault networks will in the long-term improve present attempts to develop better space-time models of earthquake triggering, which still lack information on fault localization by assuming diffuse seismicity unrelated to faults or assume very simplified structures [*Gerstenberger et al.*, 2005; *Ogata and Zhuang*, 2006; *Woessner et al.*, 2010]. A reliable association of earthquakes and faults is an important constraint to determine the spatial decay of earthquakes in aftershock sequences, which provides insights into the triggering mechanisms of earthquakes [*Felzer and Brodsky*, 2006] and improves estimates of where aftershock hypocenters are located in comparison to the main shock properties [*Hauksson*, 2010; *Powers and Jordan*, 2010; *Woessner et al.*, 2006].

Earthquake forecasting must issue statements about the likely spatial location of upcoming events. In an ideal case, we would like to forecast the set of faults or fault segments about to break in the near future. This would help predicting the expected ground motions due to radiated seismic waves, as well as anticipating problems due to surface faulting prone to cause damage on infrastructures. This goal is addressed with current fault-based approaches that use catalogs of mapped faults such as the Community Fault Model (CFM) in Southern California; (see *Plesch et al.* [2007]), which however lack the small scale structures that may contribute significantly to short and intermediate-term hazards. Moreover, as illustrated by the $M_w$ 6.7 Northridge, 1994 earthquake, a significant number of large earthquakes continue to occur on faults that were not yet mapped and were only revealed by the earthquake itself. In the case of Northern California, most of the seismicity remains unexplained by the set of mapped faults as shown for example in *Wesson et al.* [2003], where most events are labeled under 'BKGD', for 'background', whereas they seem to occur on well-defined fault structures. Moreover, such extensive fault catalogs do not necessarily exist in other parts of the world exposed to intense seismic hazard.

An improved knowledge of the underlying fault network may allow one to infer average slip rates on each fault at geological time scales and convert them into long-term average seismicity rates [*Gabrielov et al.*, 1996] possibly taking account of the information given by paleoseismological studies (see for example the National Seismic Hazard Mapping Program; *Frankel et al.* [2002]; the Uniform California Earthquake Rupture Forecast model; *Petersen et al.* [2007]; *Field* [2008]). This approach is used to provide long-term time-dependent or time independent forecasts.

The usual, and necessary, trick used in existing earthquake forecasting methods thus consists in smoothing the spatial structure of the earthquake catalog, in order to approximate the geological complexity of the local fault network. Only recently, forecast models were proposed that attempt to combine both seismicity and fault data sets in a common approach, yet blurring the knowledge of the fault structure by smoothing techniques [*Hiemer et al.*, 2013; *Rhoades and Stirling*, 2012]. Smoothing is performed using only the 2D set of epicenters (and not the 3D set of hypocenters), and this process always involves a set of arbitrary choices or parameters. The simplest smoothing consists in superimposing a regular grid onto the target area, thus coarse-graining the fault network at a homogeneous (and arbitrary) spatial resolution. A softer method consists in smoothing the set of declustered events with Gaussian kernels, whose bandwidths are adapted to optimize the quality of smoothing according to some metric [*Zechar and Jordan*, 2010]. In general, events are simply replaced by kernels that are added up over the whole space and normalized so that the integral of the spatial density of events is equal to the number of events in the catalog. In many implementations, a smoothing is considered as optimal when it maximizes the score of the forecasts on an independent dataset. It follows that the smoothing parameters do not stem from independent geological or physical knowledge. They thus look more like hidden parameters of the forecasting technique as a whole. Moreover, the use of square cells or isotropic kernels is totally opposite to what could be expected to best approximate a set of plane segments, whose orientations vary in space (see for example *Gaillot et al.* [2002] and *Courjault-Radé et al.* [2009], for the spatial analysis of sets of epicenters using anisotropic wavelets, inspired by a methodology initially developed by *Ouillon et al.* [1996]; *Ouillon et al.* [1995] for maps of fault or joint traces). In some cases, the bandwidth of the kernels may also depend on the size of the local events or on their spatial density: the larger the latter, the finer the resolution.

The well-documented multiscale organization of earthquakes and faults precludes any objective choice of the most appropriate spatial resolution to study their dynamics. The only characteristic scales in such systems are the size of the system itself (at large scales), and the scale below which scale invariance breaks down without producing bonus information; typically, this is the smallest distance between pairs of events, or the size of the smallest fault, or the width of geological and rheologically different layers [*Ouillon et al.*, 1996]. From a statistical physics point of view, one may argue that taking account of the numerous 'microscopic' spatial details of the seismicity process may only deteriorate our ability to model their dynamics and provide efficient forecasts, which is then a good reason to perform a smoothing. Another obvious reason is that events are always spatially located up to some finite uncertainties. However, *Werner et al.* [2011] brought into the debate new interesting elements by noticing that taking account of small magnitude earthquakes (down to M = 2) in the input data set increased the likelihood of the forecasts. As increasing the number of small-scale earthquakes allows one to take account of smaller-scale details of the fault network, it follows that the smoothed seismicity rate of *Werner et al.* [2011] closely reflects the best possible approximation of the fault network they could hope to get. This result echoes the conclusion of *Zechar and Jordan* [2008] who suggest that future seismicity-based techniques should also use the set of faults as a data input.

No independent and accurate geophysical technique exists that provides a detailed and complete 3D map of active fault networks. As a consequence, we rely on seismicity itself as the best proxy to image the current fault network. Continuous and recent progresses in earthquake location techniques now allow the manipulation of rather precise spatial data. For example, as absolute locations used to feature uncertainties of the order of a few kilometers in Southern California are now re-estimated using relative location algorithms, the (relative) uncertainties are now shrinking down to only a few tens of meters [*Hauksson et al.*, 2012; *Waldhauser and Schaff*, 2008]. Nonlinear location algorithms [*S. Husen et al.*, 2007; *S. Husen et al.*, 2003; *Lomax et al.*, 2009] even allow the direct sampling of the full probability density function (hereafter pdf) of the location of each event. It follows that seismologists now have the opportunity to access to the detailed topology of the active part of the fault network, provided they have the tools to estimate the position, size and orientation of fault segments from the precise location of events listed in earthquake catalogs, i.e. to extract the full value from these golden data.

*Ouillon et al.* [2008] recently proposed a new method of pattern recognition that reconstructs the active part of a fault network from the spatial location of earthquake hypocenters. It is inspired from the seminal k-means method [*MacQueen*, 1967], which partitions a given dataset into a set of (*a priori* isotropic) clusters by minimizing the global variance of the partition. *Ouillon et al.* [2008] generalized this method to the anisotropic case with a new algorithm, which, in a nutshell, fits the spatial structure of the set of events with a set of finite-size plane segments. The number of segments used is increased until the residuals of the fit become comparable to the average hypocenters location uncertainty. One can then estimate the position, size and orientation of each plane segment. *Ouillon et al.* [2008] applied this algorithm to synthetic datasets as well as to the aftershock cloud of the Landers, 1992 event, in Southern California, for which they showed that 16 planes were necessary to provide a fit compatible with the average location errors. Moreover, extrapolating the set of plane segments to the free surface, the predicted fault traces showed a good agreement with observed fault traces of the Southern California Community Fault Model (CFM) and also allowed to map faults of significant size that are not reported in the CFM.

The main shortcoming of the *Ouillon et al.* [2008] clustering method is its rough account of location uncertainties, assumed to be constant for the whole catalog. In this paper, we improve on this method by taking account of the detailed and individual location uncertainties of each event, which control both the fit through the use of the Expected Squared Distance between an event and a plane and the resolution at which the latter is performed. As the fitting method is still strongly nonlinear, different runs generally converge towards different local minima of the residuals. We thus introduce new methodologies to validate the obtained solutions, as systematic and automatic comparison with existing fault maps, if existent, is a very difficult exercise, in particular because it lacks a precise metric. We thus present six validation schemes: two of them based on the residuals of the fit, and four others based on the compatibility of the fault networks with known focal mechanisms. The new method is then tested on simple and more complex synthetic fault networks, as well as on a new catalog of the Landers area.

## 2 The optimal anisotropic data clustering (OADC) method

The new clustering method proposed here is based on a pattern recognition technique called k-means, shortly described in *Ouillon et al.* [2008] and in more details in *Bishop* [2006], *Duda et al.* [2001] and *MacQueen* [1967]. This technique makes no assumption about the shape of the individual clusters. In that sense, it can be viewed as an 'isotropic' processing of data. When dealing with earthquakes, it is desirable to cluster data within structures that can be identified as faults. In that case, the minimum *a priori* information that may help to constrain the pattern recognition process is that the clusters we look for should be highly anisotropic, i.e. that their thicknesses should be very small compared to their other dimensions.

The OADC method of *Ouillon et al.* [2008] provides an attempt to reconstruct fault networks using solely the information contained within seismicity catalogs. Compared with other strategies, e.g. the Community Fault Model (CFM) of the Southern California Earthquake Center (SCEC), it defines a general method that can identify active fault segments without taking into account direct observations such as maps of fault surface traces and/or subsurface borehole data, nor indirect observations like seismic reflection profiles to map deeper structures. *Ouillon et al.* [2008] also provide a discussion of other seismicity clustering techniques.

The OADC method is directly inspired from the original definition of the k-means method, yet generalizes it to strongly anisotropic clusters, whose thicknesses are assumed to be very small. Each fault segment is thus approximated by a finite rectangular plane, characterized by its dimension (length and width), orientation (strike and dip) and position of its center. Earthquakes are handled as pure data points, while a uniform and isotropic location uncertainty $\varepsilon$ is assumed to hold for all events.

The general algorithm of the method is the following:

1. Initialize $N_0$ planes with randomly chosen center positions, orientations and dimensions.
2. For each earthquake $\vec{O}$ in the catalog, compute the distance from it to each plane $\vec{C}$, determine the closest plane, and associate the former to the latter. Earthquake locations are treated as points, and Euclidean distances to the finite planes are computed. This first partition provides us a set of $N_0$ clusters of events.
3. For each cluster, perform a spatial principal component analysis (PCA), and use the eigenvalues and eigenvectors to define their new dimensions, orientations, and center positions. The thickness of each cluster is given by the square root of the smallest eigenvalue. The two other eigenvalues provide the length and width of the cluster (see *Ouillon et al.* [2008] for details).
4. Assuming a uniform catalog spatial location uncertainty $\varepsilon$, the computation stops if the thickness of each cluster is smaller than $\varepsilon$, as the dispersion of events across each plane can be fully explained by location errors. If there is at least one cluster for which the thickness is larger than $\varepsilon$, then proceed to step 5.
5. Split randomly the plane associated to the thickest cluster into m sub-planes, increase $N_0$ accordingly by m-1, and go back to step 2.

This procedure, which is nothing but a nonlinear fitting technique, ensures that events will be partitioned into clusters with negligible thickness (up to location uncertainties), i.e. plane-like structures, which are the assumed *a priori* model for faults.

Similarly to the classical k-means method, the OADC method may converge to a local minimum of the global clusters fit residual. One can solve this problem by running the clustering procedure several times, with different initial conditions, in order to explore the solution space and select the fault network model that achieves a genuine global minimum. However, as the method itself ensures that all fit residuals are smaller than location uncertainties, all solutions are therefore statistically equivalent. Picking one of them as the best one thus requires an independent validation process. Due to computational limitations, *Ouillon et al.* [2008] provided only ten runs on the Landers aftershocks dataset, yet noticed that the method converged more often to one of the solutions than to any other (thus suggesting a validation based on the most frequently selected solution). For each solution, extrapolating all the planes they obtained to the free surface, thus generating the corresponding predicted surface fault traces, they noticed that the most frequent solution was also the one that fits best the observed natural fault traces in this area. While offering a validation procedure on an independent dataset, this approach would prove cumbersome when dealing with much larger areas, or with zones where no such fault traces maps or incomplete ones are available. Another drawback is the subjectivity of the comparison, which is not based on any quantitative metric. The systematic validation of the obtained solutions is thus still an open problem.

Another obvious limitation of the OADC method is the assumption made about location uncertainties, which are considered to be uniform and isotropic. This hypothesis is unrealistic since focal depth is often less well constrained than the epicentral location. Moreover, location uncertainty is strongly influenced by the velocity model error, the quality of waveform pickings, the station network geometry, etc., and is thus very heterogeneous in space and time (e.g. *Stephan Husen and Hardebeck* [2010]). It thus follows that the clustering process should be more detailed in some areas and sparser in some others. The clustering method should take this heterogeneity into account.

# 3 Anisotropic clustering of location uncertainty distributions (ACLUD)

The original k-means method assumes that the uncertainty of the spatial location of data points is negligible. In the case of real physical systems, the story is different. For earthquakes, location uncertainty is an inherent property due to wave arrival time inaccuracy, velocity model errors, station network geometry, or outdated data sources like historical seismicity catalogs. When taking uncertainty into account, data can no longer be described as a point-process, but as a more or less complex probability density function (hereafter pdf). *Chau et al.* [2006] claim that uncertainties can significantly affect the results provided by clustering techniques such as k-means. They thus introduce the *uk-means* algorithm (where 'u' stands for 'uncertain', see electronic supplement), which incorporates uncertainty information and provides, when considering synthetic samples, more satisfying results than the standard algorithm.

We now show how to extend the uk-means method of *Chau et al.* [2006] to the case where the cluster model $\vec{C}$ is a plane, in the spirit of *Ouillon et al.* [2008], and the object to cluster $\vec{O}$ is the pdf of an earthquake location. We term the new method the "anistropic clustering of location uncertainty distributions" (ACLUD).

*Chau et al.* [2006] suggest using the expected squared distance (hereafter ESD), which, in our case, is defined as:

$$d^2(\vec{O},\vec{C}) = \int_{\vec{x}\in\vec{O}} \left\|\vec{x}-\vec{C}\right\|^2 f(\vec{x})d\vec{x}$$
$$= \int_{\vec{x}\in\vec{O}} \inf_{\vec{c}\in\vec{C}}\left\|\vec{x}-\vec{c}\right\|^2 f(\vec{x})d\vec{x} \qquad (1)$$

where f(x) is the pdf of the earthquake location. While this distance is easily estimated in the case of an infinite plane $\vec{C}$, we also propose computationally efficient approximations in the case of a finite-size plane.

## 3.1 Expected square distance (ESD) between a probability density function and an infinite plane

We consider an infinite plane within a Euclidean three-dimensional space. The coordinate system is chosen such that its origin is located on the plane, whose orientation is given by two of the basis vectors, the third one being normal to it. Then, Eq. (1) can be rewritten as:

$$d^2(\vec{O},\vec{C}) = \int_{\vec{x} \in \vec{O}} \|x_3\|^2 f(\vec{x}) d\vec{x} \quad (2)$$

where $x_3$ is the third component of point $\vec{x} \in \vec{O}$. Noticing that:

$$x_3^2 = (x_3 - k_3 + k_3)^2 = (x_3 - k_3)^2 + k_3^2 + 2(x_3 - k_3) \cdot k_3 \quad (3)$$

with $k_3$ being the third component of the centroid of $\vec{O}$, and given that the contribution of the last right-hand term of Eq. (3) to the integral is zero, Eq. (2) becomes:

$$d^2(\vec{O},\vec{C}) = k_3^2 + \int_{\vec{x} \in \vec{O}} (x_3 - k_3)^2 f(\vec{x}) d\vec{x} \quad (4)$$

The first term in the right-hand side is simply the squared distance between the centroid of $\vec{O}$ and the infinite plane, while the second term is simply the variance of $\vec{O}$ in the direction normal to the plane (which can be deduced from the pdf of $\vec{O}$ and its covariance matrix). This is nothing but the variance decomposition theorem.

## 3.2 Expected square distance (ESD) between a probability density function and an infinite line or a point

Following a similar procedure when $\vec{C}$ is a line, we can choose a coordinate system so that $\vec{C}$ lies on the first axis. Then we get:

$$d^2(\vec{O},\vec{C}) = k_2^2 + k_3^2 + \int_{\vec{x} \in \vec{O}} \left[(x_2 - k_2)^2 + (x_3 - k_3)^2\right] f(\vec{x}) d\vec{x}$$
$$= \sum_{i=2}^{3} k_i^2 + \sum_{i=2}^{3} \int_{\vec{x} \in \vec{O}} (x_i - k_i)^2 f(\vec{x}) d\vec{x} \quad (5)$$

When $\vec{C}$ is a point, we can choose a coordinate system so that $\vec{C}$ lies at the origin. Then we get:

$$d^2(\vec{O},\vec{C}) = \sum_{i=1}^{3} k_i^2 + \sum_{i=1}^{3} \int_{\vec{x} \in \vec{O}} (x_i - k_i)^2 f(\vec{x}) d\vec{x} \quad (6)$$

The interpretation of Eq. (5) and (6) is the same as for Eq. (4) except that we now compute the distance between the centroid and a line and use the relevant dimension for the variance decomposition. This last set of equations will prove very useful when approximating the distance between a pdf and a finite plane.

## 3.3 Expected square distance (ESD) between a probability density function (pdf) and a finite plane

The anisotropic clustering of location uncertainty distributions (ACLUD) method we propose still assumes that active fault segments can be modeled as rectangular finite planes. If it proves rather easy to compute the Euclidean distance between a point and a finite plane, the problem is a bit more difficult when observations are given through their pdf's. Indeed, we shall see that, using the variance decomposition theorem, we can only provide theoretical approximations to the expected squared distance between a pdf and a finite plane.

Figure 1 illustrates the problem. The grey rectangle area represents a finite plane $\vec{C}$. Events may be located anywhere in the full 3D space that surrounds it. We now consider any object $\vec{O}$ in the 3D space and its projection $\vec{Q}$ along the direction normal to $\vec{C}$ onto the infinite plane containing $\vec{C}$. The object $\vec{Q}$ will be located within one or more of the nine sectors defined in Figure 1, each sector being indexed in roman numbers from I to III as shown in the figure. The object $\vec{Q}$ can overlap several sectors depending on the shape of the support of its pdf.

If $\vec{Q}$ is completely included within sector III, then the ESD between $\vec{O}$ and $\vec{C}$ can be computed using Eq. (4), as the infinite plane assumption is valid. If $\overline{Q}$ is completely included within a sector labeled (II), the ESD is computed using Eq. (5) (after an appropriate change of coordinates) as the infinite line assumption is valid. If $\overline{Q}$ is completely included within a sector labeled (I), the ESD should be computed between the pdf and the closest corner of the finite plane, using Eq. (6). Indeed, a similar approach has been used in *Ouillon et al.* [2008] to compute the Euclidean distance between a given hypocenter and a given finite plane.

In our case, the general problem is much more complex as we implicitly have to consider the distance between the finite plane and every point where the pdf of $\overline{O}$ is defined. This implies that the projection $\overline{Q}$ is characterized by a pdf that may overlap several distinct sectors, so that none of the above simple formulae (4), (5) and (6) can be used anymore. In that case, only a direct Monte Carlo approach provides an accurate estimate of the ESD. As it would prove computationally too heavy when handling large catalogs and sets of faults, we propose a simplification: we first consider only the centroid of $\overline{O}$ and its own projection. If the latter is contained within sector III, we use formula (4) as an approximation to the ESD. If it is contained within a sector labeled (II), we use formula (5). If it is contained within a sector labeled (I), we use formula (6). This approximation is obviously wrong when the size of the finite plane is much smaller than the spatial extent of the domain where the pdf of $\overline{Q}$ is defined. However, in practice we found that for most of cases, location uncertainties are much smaller than the size of potential fitting fault plane we can resolve.

### 3.4 Anisotropic clustering of location uncertainty distributions algorithm

Assume that an earthquake catalog provides the location of each event with a pdf. We can characterize the location with its centroid (hereafter, the hypocenter) and its covariance matrix. The new clustering algorithm we propose is the following:

1. Split randomly the earthquake catalog into 2 distinct subsets: the training set (which is the one to be fitted) and the validation set (which is the one used to qualify or discriminate different clustering models).
2. Initialize a number of $N_0$ faults with random positions, orientations and dimensions.
3. For each earthquake in the training subset, associate the earthquake to the closest plane according to the ESD. We thus get a partition of events into a set of $N_0$ clusters.
4. For each cluster $i$, compute the covariance matrix of the locations of its associated hypocenters, and find its eigenvalues and eigenvectors. By doing so, the dimensions and orientations of each cluster can be computed. The smallest eigenvalue $\lambda_{i,3}$ provides the thickness of the corresponding cluster.
5. For each cluster, compute the average individual variance $\varepsilon_i$ of the hypocenters' location pdf in the direction normal to the cluster.
6. For each cluster, compare its thickness $\lambda_{i,3}$ with the average location uncertainty $\varepsilon_i$ of its associated events. If $\varepsilon_i \geq \lambda_{i,3}$ for all clusters, the computation stops, as location errors alone can explain the finite thickness of each cluster. We then proceed to step 8. If there is at least one cluster for which $\varepsilon_i < \lambda_{i,3}$, then we proceed to step 7 as we need more planes to explain the data.
7. We split randomly the thickest cluster into m other planes, and go back to step 3 (increasing $N_0$ accordingly by m-1).
8. We compute the residual of the fit of the validation data set conditioned on the fault network model of the training data set (from step 6).
9. We repeat steps 1-8 many times (typically several thousands) and rank all models according to their validation fit residuals obtained in step 8.

For this study, in step 7 we use m=2. The proposed algorithm accounts for individual event location uncertainties, both in the computation of the ESD between an event and the planes and in the criterion used to continue or stop the fitting process. The stopping criterion thus doesn't assume a spatially uniform location uncertainty, but is adapted to the case of space-dependent location quality. This property is particularly welcome in the case of earthquakes for which location uncertainties heavily depend on the spatial structure of the stations networks.

One should also be aware that the full three-dimensional confidence interval is different from the confidence interval in 1D. In order to compute the variance of the pdf in the direction normal to the plane, we have to project the 68% three-dimensional confidence ellipsoid onto that normal direction. Yet, after the projection, the confidence level increases to higher levels so that the correct quantiles have to be estimated (*Press et al.* [2007] , page 811, figure 15.6.3).

By subdividing the data set, we implement a cross-validation technique to the predictive skill of the clustering approach. Our procedure separates randomly the full dataset into two independent subsets, generates the fault model that fits the training dataset and evaluates it by estimating how well it predicts the independent validation set. The process is repeated several times, each trial corresponding to different training and validation sets, and we select the one with the best validation result. How to generate the training and validation data sets is a question in itself. On the one hand, if there are not enough earthquakes in the training set, it will lead to a spurious fit with a very bad validation score; on the other hand, if there are not enough earthquakes in the validation set, residuals may fluctuate and depend strongly on the particular choice of the validation set. From synthetic tests where the original fault networks are known, we checked that using 95% as training and the remaining 5% as validation data generally provides robust results.

The main assumption of this algorithm is that the hypocenter corresponds to the expectation hypocenter location [*Lomax et al.*, 2000]. In the framework of probabilistic earthquake location the hypocenter location is usually associated with the maximum likelihood point [*Tarantola and Valette*, 1982]. The assumption that the hypocenter is not very different from the maximum likelihood point would be valid if and only if the pdf of the location of the event is compact, i.e. small in size, which has no *a priori* reason to be true. We shall discuss later the conditions for which this assumption might be approximately valid in the case of natural earthquakes catalogs.

## 3.5 Validation strategies

The new clustering method automatically explores a very large solution space. In order to find the "best" solution, we follow a purely statistical strategy, i.e. cross validation. However, other validation strategies might be more appropriate. In the following, we will introduce three other criteria: one residuals-based statistical strategy called Bayesian Information Criterion (BIC, see *Schwarz* [1978]), and two metrics based on observed focal mechanisms.

### 3.5.1 Bayesian Information Criterion

BIC is a commonly used statistical criterion for model selection that takes both the likelihood function and model complexity into consideration. During clustering, it is possible to increase the likelihood by adding more faults, at the cost of increasing the complexity of the model. By adding a penalty term for the number of faults, the BIC merges the likelihood and complexity of the solution together. Assuming that the distribution of earthquakes across the fitting planes is a normal distribution, the BIC can be expressed as:

$$BIC = n \cdot ln(\hat{\sigma}) + k \cdot ln(n) \qquad (7)$$

where *n* is the number of events used for the fit, $\hat{\sigma}$ is the unbiased variance estimation of the earthquake distribution across the fitting planes, and *k* is the number of faults in the tested model. Thus, by minimizing the BIC, we may find the best network from the solution space that provides both a large likelihood and a simple model structure. The difference with the cross validation scheme is that the latter is performed using the validation dataset, whereas the BIC uses the training dataset. It is also important to notice that, during the clustering process, we randomly partition the whole data set into training and validation sets. It means that, for each clustering run, the training set changes so that the computed BIC is not strictly derived from the same training set. However, considering that we deal with large datasets among which 95% of each single one is used as training sets, the BIC remains a robust estimator.

### 3.5.2 Focal mechanism µ-metrics

The focal mechanism of an earthquake describes the potential orientations of the rupture plane and slip vector. If events are clustered together on a given fault plane, we may expect them to be characterized by similar focal mechanisms, the latter being also consistent with the orientation of the fitting plane. This provides a mechanical approach to validation. At the end of each fit, we thus adopt the following procedure:

1. For each cluster, select the available focal mechanisms of events.
2. For each focal mechanism, compute the normal vector to each of the two nodal planes.
3. For each nodal normal vector, compute its dot product with the vector normal to the cluster (defined as pointing upwards). If one of the dot products is negative, replace the nodal normal vector by its opposite and change the sign of the dot product.
4. From both nodal normal vectors, choose the one that maximizes the dot product.
5. Once steps (2)-(4) have been fulfilled for each event of the cluster, stack all the selected nodal normal vectors, and compute the angle µ between the resultant and the normal vector to the cluster.

Step (5) is performed after weighting each selected nodal normal vector according to the magnitude M of the corresponding event. The weight is taken as $10^{a \cdot M}$. If $a=0$, then all events have the same weight and the measured angular discrepancy is mainly controlled by the smallest events. If $a=1/2$, then each event is weighted proportionally to its empirically assumed slip amount, while it is weighted by its energy or moment if we set $a=3/2$ (and in that case the angular discrepancy is controlled by the largest event in the cluster). Our choice is to consider the case $a=1$, as each event is weighted proportionally to its empirically assumed rupture area. Moreover, if the local Gutenberg-Richter b-value is close to 1, each magnitude range contributes equally to the estimated angular discrepancy.

We first define a weighted average normal vector to the selected nodal plane of events on fault plane $F_i$ as:

$$\vec{V}_{E_i} = \frac{1}{\left\| \sum_{k=1}^{m(i)} \vec{v}_{E_{i,k}} \cdot 10^{a \cdot M_{i,k}} \right\|} \sum_{k=1}^{m(i)} \vec{v}_{E_{i,k}} \cdot 10^{a \cdot M_{i,k}} \qquad (8)$$

where:

- $m(i)$ = the number of events in fault plane $F_i$;
- $\vec{v}_{E_{i,\cdot}}$ = the normal vector to the selected nodal plane of a given event on fault plane $F_i$;

We then define a global angular discrepancy of the full set of planes as the $\mu_{fault}$ measure. It is formally expressed as:

$$\mu_{fault} = \frac{\sum_{i=1}^{n} cos^{-1} \left| \vec{V}_{F_i} \cdot \vec{V}_{E_i} \right|}{n} \qquad (9)$$

where:

- $n$ = the number of fault planes;
- $\vec{V}_{F_i}$ = the normal vector to fault plane $F_i$;

The weighting strategy of Eq. (9) implies that we simply compute an average angular misfit over all faults (hence the associated subscript on the left-hand side). Similarly, we can also perform the average over all events. We obtain:

$$\mu_{event} = \frac{\sum_{i=1}^{n} m(i) cos^{-1} \left| \vec{V}_{F_i} \cdot \vec{V}_{E_i} \right|}{\sum_{i=1}^{n} m(i)} \qquad (10)$$

Minimizing both estimators will select networks where the orientation of inverted fault planes is the closest to the average orientation of the focal mechanisms. In summary, the µ-metric measures the magnitude weighted average direction of the normal vectors of the "observed" focal mechanisms to the normal vector of the fault plane derived within the clustering approach.

### 3.5.3 Focal mechanism σ-metrics

Events grouped together by our fitting procedure may also feature roughly similar focal mechanisms, whose orientation may be different from the one of the fitting plane (see sketch in Figure 2 and explanations below). Following the same procedure as above from step (1) to (4), we change step (5) as:

6. Once steps (2)-(4) have been fulfilled for each event, stack all the selected nodal normal vectors, and compute the average angle between each individual selected nodal vector and the resultant stacked vector.

The associated measures are defined $\sigma_{fault}$ and $\sigma_{event}$, depending on the way they are averaged. They are similar to standard deviation in statistics, yet we compute them using the L-1 norm (and not the L-2 norm). The reason is that, in the case when the distribution of angles is not Gaussian but fatter tailed, using the L-1 norm provides results less sensitive to large outliers. Using the same notations as above, the mathematical expressions are:

$$\sigma_{fault} = \frac{\sum_{i=1}^{n} \frac{1}{m(i)} \sum_{j=1}^{m(i)} cos^{-1} \left| \vec{v}_{E_{i,j}} \cdot \vec{V}_{E_i} \right|}{n} \quad (11)$$

$$\sigma_{event} = \frac{\sum_{i=1}^{n} \sum_{j=1}^{m(i)} cos^{-1} \left| \vec{v}_{E_{i,j}} \cdot \vec{V}_{E_i} \right|}{\sum_{i=1}^{n} m(i)} \quad (12)$$

σ measures the angular difference from each single normal vector to the fault plane from the clustering approach and then averages, which results in a quite different metric.

Figure 2 shows examples of applying µ and σ measures. On each plot, the black line indicates the trend of the fault zone, while the gray lines indicate the potential orientations of shorter individual ruptures within the fault zone, all events being clustered in the same macroscopic fault zone (see Section 6 for a further discussion of the influence of the fault zone complexity on the results of clustering). When the rupture planes are quasi-colinear with the fault trend, then both µ and σ values are small (Fig. 2a). Figure 2b shows a series of planes for which orientations oscillate around the trend of the fault. In that case, the µ value is still small while the σ value is larger. Figure 2c shows the case of an *en-échelon* distribution of rupture segments, which will provide a finite and possibly large µ value and a very small σ value. The last example (Fig. 2d) shows a series of alternating conjugate rupture planes, which will be associated with large values of both µ and σ. These two measures derived from focal mechanisms can quantify the degree of agreement of the reconstructed fault network with local focal mechanisms. They provide tools in model selection with consideration of tectonic knowledge compared to pure statistical approaches such as cross-validation or BIC.

## 4 Tests of the ACLUD method on synthetic catalogs featuring location uncertainties

The previous section has introduced a new clustering scheme to automatically reconstruct fault structures from seismicity catalogs including location uncertainty information. We apply the approach to synthetic catalogs to understand its sensitivity to different structural complexities.

### 4.1 Generation of datasets

Locating earthquakes results in a posterior probability density function of an event location [*Moser et al.*, 1992; *Tarantola and Valette*, 1982; *Wittlinger et al.*, 1993]. The pdf may possess any arbitrary shape and may be visualized using scatter density plots, which are obtained by drawing samples from the posterior pdf with their number being proportional to the probability [*S. Husen et al.*, 2003; *Lomax et al.*, 2000]. From these samples, the 68% confidence ellipsoid can be computed by a singular value decomposition of the corresponding covariance matrix, and consists in a rough approximation of the spatial uncertainty of the location estimate. The expectation hypocenter is at the center of the confidence ellipsoid, and the maximum likelihood hypocenter will always be located within the densest part of the pdf, so that both locations do not necessarily coincide.

In this section, we generate synthetic earthquake catalogs using the NonLinLoc software package (*Lomax et al.* [2000], Version 5.2, http://alomax.free.fr/nlloc/). Compared to traditional, linearized approaches, NonLinLoc is superior in that it computes the posterior pdf using nonlinear, global searching techniques. The general method we use to generate a synthetic earthquake catalog is the following. We first impose the geometry of the original fault network, which consists in a collection of rectangular planes with variable locations, sizes and orientations. We then assume that all earthquakes occur exactly on those planes and generate P-waves. We then randomly distribute a given number of earthquakes on those planes. For a given velocity model, theoretical travel times between the true hypocenters and a set of given stations are computed. Random perturbations are added to the arrival times mimicking the uncertainty in picking waveform onset, which allows us to proceed to the inverse problem of computing the location of the events as well as their uncertainties by using NonLinLoc.

To generate the set of associated synthetic focal mechanisms, we first assume that the rake of the slip vector on each plane is zero. For each event, the strike and dip are assumed to be identical to the ones of the input plane to which it belongs. We then add an independent Gaussian random perturbation respectively to the strike, dip and rake of the event. Those perturbed angles are then used to compute the strike and dip of the auxiliary plane, thus providing a complete focal mechanism.

Note that we did not take account of the possible errors on the velocity model, which would provide systematic errors on both locations and focal mechanisms.

The catalog of relocated hypocenter locations including their scatter density clouds is then fitted with a set of finite planes, using the ACLUD algorithm as defined in the previous section. The best solution which depends on the validation technique is then compared to the original input fault network.

As a first test, we generated a very simple synthetic dataset consisting in three vertical faults featuring 4,000 events in all (thus similar to the one studied in *Ouillon et al.* [2008]) and characterized by their full pdf. The new clustering technique we propose successfully reconstructed the fault network whatever the validation criterion we used (see electronic supplement). We shall now test it on a more realistic and complex case.

## 4.2 Synthetic catalog with complex geometry inspired from Ouillon et al.

This synthetic dataset outlines a more complex and realistic case. Figure 3a shows the structure of the reconstructed fault network in the area of the 1992 $M_W$ 7.3 Landers earthquake by *Ouillon et al.* [2008]. It features 13 planes with a dip larger than 45° (the three other planes, dipping less than 45°, have been removed as they certainly are spurious planes – see *Ouillon et al.* [2008]). The original catalog used in *Ouillon et al.* [2008] includes 3,103 events, which we now assume to occur randomly and uniformly on those planes. We define a virtual station network, similar to the simpler one used in the example shown in the electronic supplement, in order to compute theoretical wave travel times to 11 randomly chosen stations, and add Gaussian errors with a standard deviation of 0.1 s to simulate picking errors. Figure 3b shows the spatial distribution of the relocated 3,103 events. To generate the set of synthetic focal mechanisms, we add an independent Gaussian random perturbation respectively to the strike, dip and rake of each event with a standard deviation of 10°. Those perturbed angles are then used to compute the strike and dip of the auxiliary plane, thus providing a complete focal mechanism. Note that magnitude is not taken into consideration, so that we attribute the same magnitude to all events.

Note that, in Section 3.5, we defined two statistical measures derived from focal mechanisms, which can be used to evaluate each reconstructed fault network. We can also assess the individual contribution of each cluster with respect to those global measures. We then similarly define for each cluster two individual measures of focal mechanism consistency, $\mu_F$ and $\sigma_F$:

$$\mu_F = cos^{-1}\left|\vec{V}_F \cdot \vec{V}_E\right| \qquad (13)$$

$$\sigma_F = \frac{1}{m}\sum_{i=1}^{m} cos^{-1}\left|\vec{v}_{E_i} \cdot \vec{V}_E\right| \qquad (14)$$

where:

- $m$ = number of events within fault $F$;
- $\vec{V}_{F_i}$ = normal vector to the given fault plane $F$;
- $\vec{V}_E$ = weighted average normal vector to the selected nodal plane of events on fault $F$;
- $\vec{v}_{E_i}$ = normal vector to the selected nodal plane of a given event on fault $F$;
- $M_i$ = magnitude of a given event.

A large $\mu_F$ value indicates that the average focal mechanism rupture plane deviates significantly from the fitted fault plane. A large $\sigma_F$ value indicates a significant dispersion of the orientations of focal mechanisms within the cluster.

We performed 6,000 runs with different initial conditions of the random number generator which controls the fault splitting step, and obtained as many solutions. We now discuss the results obtained using the six validation techniques discussed in Section 3.5.

**Cross validation:** Figure 4a shows the selected reconstructed network, featuring 14 planes. One can notice that two faults in the northern end are merged into a single plane. This is due to the fact that locations quality in this region is deteriorated due to a poor station coverage. Such a poor coverage also occurs for the southern end, where the two crossing faults are reconstructed as a set of three faults. This kind of local overfitting is often observed in such situations, and is due to the splitting step of the clustering process.

**BIC:** Figure 4b shows the selected reconstructed network, featuring 15 planes, which is different from the one selected by cross-validation. Whereas the structure is now correctly inverted in the northern part, one can observe a small fault in the middle region pointed by the arrow, whose orientation is clearly rotated clockwise compared to the original synthetic network (Figure 3a). The reason is that the BIC gives more weight to the fault planes featuring more events. The density of events on this fault is the smallest among all 15 faults (for which this parameter ranges from 0.6/km$^2$ to 5.0/km$^2$). The reconstruction of such low event density faults can be unstable as their weight in the global criterion is very small. We also noticed that its individual $\sigma_F$ value is the largest (for which this parameter ranges from 12° to 29°), indicating that the focal mechanisms of events clustered on this fault are very scattered.

**$\mu$ metrics:** both $\mu_{event}$ and $\mu_{fault}$ metrics select the same solution, shown in Figure 4c, featuring 14 planes. One can observe that two faults in the northern middle region have been merged into a single one (indicated by a small arrow). The distribution of $\mu_F$ and $\sigma_F$ values of all 14 planes range from 1° to 43° and 12° to 23°, respectively. The individual $\mu_F$ value and $\sigma_F$ value of this merged fault are both the largest over all 14 faults. This indicates that the focal mechanisms of the events clustered on this fault are neither consistent with each other nor with the orientation of the fitting plane. This thus makes the fault suspicious. More runs would be necessary to sufficiently sample the solution space and get a fully correct solution.

**$\sigma$ metrics:** Figure 4d shows the reconstructed network chosen by both $\sigma_{event}$ and $\sigma_{fault}$ metrics, featuring 13 planes. Three faults in the central region are merged into a single large fault (see the arrow). This comes from the fact that the orientations of those three faults are very similar. The individual $\mu_F$ and $\sigma_F$ values of this merged fault are close to the average of the values obtained on the other planes. We thus have no way to diagnose this cluster as abnormal. This may stem from the fact that the faults that generate those events are located close to each other and feature orientation differences less that the uncertainties on the focal mechanisms orientations.

Figure 5 shows the stereo plots of the original input faults and of the four solutions favored by the six different criteria. Plots in the left column indicate the orientations of fault traces. Dots in the right column indicate the directions of the normal vectors to the fault planes. Qualitatively, there is a nice agreement between all the reconstructed networks and the true network (first row of Figure 5).

This little example shows that inverting a complex but realistic structure, given realistic location uncertainty estimates, is not an easy task. However, the inverted networks, if not identical to the original one, are very similar to the original synthetic ones using the selection criteria. All validation criteria feature reasonable solutions: none of them is particularly better or worse than any of the others and the selections based on pure statistical techniques give similar fault networks as those based on tectonic constraints.

## 4.3 Comparison of the ACLUD method with the OADC method.

The OADC method uses a single, uniform and isotropic location uncertainty for the whole catalog as the clustering stopping criterion. For the synthetic Landers catalog, we computed an average location uncertainty of 1.10 km. Using this value as the stopping threshold, we performed 6,000 runs using the code of *Ouillon et al.* [2008]. The OADC method does not feature the cross validation procedure, so that all events are used as the training data set. However, we can still rank all 6,000 solutions based on their final clustering global residuals. Figure 6 shows the four solutions chosen by the six following criteria: best global clustering residual, BIC, and the four focal mechanism criteria previously defined. All those four solutions selected from different criteria clearly miss the small-scale structure of the network. Obviously, clustering has been forced to end too early due to using an inappropriate average location uncertainty estimate, especially in the central region. As location uncertainties in the central region are smaller than close to the northern and southern edges (due to a better station coverage), the stopping criterion, that resembles the location uncertainty, should be smaller in the central region than in the edge regions. Clustering thus stopped too early in the central region and made the structure coarser. Comparing with our new method, we thus clearly see the advantage of using the true location uncertainty of each event. Comparing the four solutions, we notice that the three of them chosen by focal mechanism criteria are superior to the ones chosen by both the global clustering residuals and the BIC criteria (see Figure 6). These three solutions cover most of the input fault planes, yet do not include planes sampled by a small numbers of events. However, despite its simplicity, the main advantage of OADC is its fast convergence.

## 4.4 Synthetic data with background events.

The previous section showed that our technique is able to reasonably reconstruct the structure of the synthetic fault network. We now test a new assumption where the catalog of events consists in the same set as before, but now we add background events. In nature, such events also occur on faults but the latter are, for our approach, undersampled by seismicity; thus a clustering technique cannot reconstruct the structure. Specifically, we add another 20% background events to the synthetic data set uniformly distributed in the 3D space (see Figure 7). The latitude, longitude and depth ranges are identical to the ones of the fault-related events, providing a total number of 3,724 events. For the sake of simplicity, their focal mechanism is chosen randomly among the set of the original 3,103 events.

Our new clustering technique follows the same approach as the OADC method to detect and remove background events. The detection is based on a local density criterion, as well as on the impossibility to associate an event with a given cluster without increasing too much its thickness. However, background events are not removed from the dataset if they are located close to a fault, as they are then undistinguishable from other events.

Results obtained using the different selection procedures are shown in Figure 8, after 6,000 runs. Both purely statistical criteria (cross validation and BIC) select models with clearly spurious faults. For example, for cross validation, we observe a large nearly horizontal plane in the northern area while, in the southern region, original planes are divided into many small planes. Similarly, for the solution selected by the BIC, a large nearly horizontal plane is generated at latitudes 46.0° - 46.2°. Those low-dipping planes are indicated by numbers on Figure 8 and Figure 9. The best results emerge when using models selected by criteria based on focal mechanisms. Looking at the properties of each cluster (see Figure 9), we notice that the reconstructed horizontal faults marked 1 and 2 have very large $\mu_F$ values. This suggests that these shallow-dipping planes disagree with their associated focal mechanisms. Results chosen by cross validation and BIC clearly show the effect of these nearby background events, which distort the inverted network and require to introduce spurious shallow-dipping faults to decrease the variance of the fit. In contrast, due to the fact that background events come mostly with arbitrary mechanisms, the validation criteria based on focal mechanisms detect more efficiently the associated inconsistencies, and favor more realistic solutions.

### 4.5 Summary of synthetic tests

The synthetic tests show that our new ACLUD method successfully reconstructs fault networks, both in the case of simple or more realistic and complex structures. The tests show that, due to location uncertainties, faults that are close in space and orientation may merge into a single structure. Comparing with the previous OADC method proposed by *Ouillon et al.* [2008], the new method improves the results by considering location uncertainties of each individual event, thus allowing us to invert the structure more finely within areas benefiting from a better station coverage. The new method also improved the validation step, as we automated the computation of six criteria, two of them being purely statistical indices of the fit (cross validation and BIC), the four others being based on the comparison between the inverted network and the observed focal mechanisms. While all those criteria provide reasonable selected models in the absence of background events, criteria based on focal mechanisms outperform the others when such background events are present. We even obtain better solutions when including background events, which may be due to a different exploration of the solution space. For real datasets, this implies performing an extensive simulation effort to reconstruct a fault network, similar to larger scale Monte-Carlo simulations. The multiple selection criteria and their characteristics also suggest that the technique does not allow us to pinpoint single best solutions but rather emphasize that possible solution groups exist, which is likely a result of undersampling of the structures with earthquakes.

## 5 Application to the Landers aftershock series

We now apply our new clustering technique to a real dataset in the area of the 1992 $M_w$ 7.3 Landers earthquake, already studied by *Ouillon et al.* [2008]; this allows for a comparison of results. The catalog we used is the one from E. Hauksson (California Institute of Technology, personal communication), which has been located using the NonLinLoc method described in the electronic supplement. The catalog contains 20 years of data from 1984 to 2004, with depth ranging from 1.37 km to 26.99 km. Unfortunately, this catalog neither features the complete description of the original pdf of event locations, nor the corresponding covariance matrices that we need to input into our clustering scheme. Uncertainty is simply characterized by the lengths and orientations of the axes of the 68% confidence ellipsoid. Note that the corresponding derivation of the covariance matrix can be rigorously achieved only when the location pdf is Gaussian, a condition which generally holds only in areas well covered by a dense network of stations [*Stephan Husen and Hardebeck*, 2010; *Lomax et al.*, 2009]. We assume that this is the case in the Landers area, due to the presence of numerous stations belonging to the permanent Southern California network, as well as due to the set of temporary stations installed during the Joshua-Tree-Landers earthquake sequence. This is also the reason why we selected a subset of events that are most likely to be located with Gaussian uncertainties, i.e. those whose locations are particularly well constrained according to the criteria we defined in a companion work (Wang et al., in preparation). We finally retained only events located using more than 11 stations, with local magnitude M≥2, and located within an area well-covered by the station network (primary azimuthal gap smaller than 180°, ratio of the epicentral distance to the closest station over focal depth smaller than 1.5, see *Bondar et al.* [2004]), yielding a final subset of 3360 events, comparable in size with our most complex synthetic example.

As the clustering technique can be considered itself partly as a stochastic process, we performed 30,000 different runs in order to reasonably sample the complex landscape of the solution space. Figure 11 shows the fault networks corresponding to the best solutions selected from the six validation procedures. Plots present the horizontal projection of the fitting plane segments, as well as the epicenters of their associated events. For the sake of clarity, the clusters obtained for each fit are split into 2 subsets depending on their dip: clusters with dip larger than 50° (left plot) and clusters with dip smaller than 50° (right plot). As the Landers area is dominated by strike slip faulting on nearly vertical faults, we think, in the spirit of *Ouillon et al.* [2008], that the large-dip clusters may represent genuine underlying faults, while the low-dip clusters mainly represent spurious structures artificially introduced in order to decrease the local residual of the fit in areas of diffuse seismicity.

Each of the validation techniques yields a different solution. Clearly, there is a large number of events that are clustered on low-dip faults (dip < 50°) in the model selected by cross-validation. Looking at the properties of each cluster, we notice that there is a clear decrease of $\mu_F$ value with increasing dip, suggesting that low-dip planes disagree with their associated focal mechanisms. Thus, the solution selected by cross validation seems not to be realistic. The other validation processes yield solutions that offer a nice agreement in the Northern part of the network (which can then be considered as reasonably well inverted), yet significant differences occur at other locations. If we leave aside the BIC solution for reasons explained in the section dealing with synthetic examples, we are left with four solutions that all agree well with focal mechanisms, and among which no definitive and objective choice can be made.

The fact that these validation techniques yield different selected solutions may come from the interplay of two main factors: the multiscale structure of individual faults and the spatial extent of earthquakes location uncertainties. Many studies show that faults feature a complex inner structure consisting of a complex subnetwork of sub-faults and secondary brittle structures [*Tchalenko*, 1970; *Tchalenko and Ambrasey*, 1970]. If the time span of the catalog is much shorter than the typical time scale necessary to activate rupture on every substructure, then most of the sub-faults will feature very few events, precluding their detailed reconstruction. Furthermore, if location uncertainties are larger than the typical spacing of sub-faults, the solution to the fit of the full network is not unique either and different validation techniques will favor different solutions.

Following the same approach as *Ouillon et al.* [2008], we also computed the predicted surface traces of the reconstructed faults for each selected model. The idea is to prolong fault planes to the surface and compare them with the observed traces compiled by the CFM (see Figure 12). None of the six predicted trace maps fully agrees with the observed surface fault traces. It may stem from the fact that the catalog we used is only 20 years long, whereas surface fault traces derive from millions of years of tectonic deformation. The active part of this network is thus necessarily a subset of the full network, so that the correspondence between both sets of fault traces is necessarily imperfect. Surprisingly, *Ouillon et al.* [2008] obtained a solution with a more realistic predicted map of fault traces in the same area.

# 6 Discussion and Conclusions

## 6.1 Summary of the results

In this paper, we introduced a new technique (the ACLUD method) to reconstruct active fault networks which improves on the method of *Ouillon et al.* [2008] as it uses both earthquake locations and their estimated individual uncertainties. After a massive, yet non-exhaustive search through the very large solution space, the full set of potential solutions is submitted to six different validation procedures in order to select the corresponding best solutions. Two of the validation steps (cross-validation and BIC) process the fit residuals, while the four others look for solutions that provide the best agreement with independently observed focal mechanisms. Tests on synthetic catalogs allowed us to qualify the performance of the fitting method and of the various validation procedures. The method is able to provide exact reconstructions in the case of very simple structures, yet is not able to find the input network when structures display more complexity and realistic location uncertainties. However, the solutions provided by each validation step are close to the expected one, especially for the BIC and focal mechanism-based techniques. Adding a uniform spatial background seismicity rate, both validation techniques based on fit residuals fail, while the ones based on focal mechanisms consistency show a much better agreement with the expected solution.

We compared the results obtained by our new ACLUD technique with the ones obtained on the same dataset using the OADC code developed by *Ouillon et al.* [2008]. Despite a slight difference in the nature of one of the validation procedures, we showed that the new method improves significantly on the OADC method, because accounting for individual location uncertainties of events allowed a more detailed fit of faults in areas where such uncertainties were small. It also showed that the results provided by the OADC method also improved when using validation steps based on focal mechanisms consistency. This last observation thus suggests the systematic use of such validation tools, whatever the underlying clustering technique. This also suggests that including focal mechanisms into the clustering scheme itself will provide a more consistent and efficient exploration of the solution space.

The technique has also been applied to a real data set, namely the Landers area. This study confirms that cross-validation provides a poor quality solution, as the network features a significant number of planes with a very low dip, at odds with the prior structural knowledge we have about the nature of faulting in that area. The obtained fault networks also show a poor agreement with focal mechanisms. Comparing the predicted map of fault traces for each of the six selected solutions to the actually observed map did not allow us to draw any conclusion. The reason why *Ouillon et al.* [2008] obtained a solution with a more realistic predicted map of fault traces in the same area remains unclear, as they did not use the same catalog. The latter may have been of lower quality than ours, which in turn allowed them to fit correctly the gross features of the network. In our case, a better assessment of locations and uncertainties may better reveal the genuine small-scale complexity of the network, which may in turn impact on the quality of the fit, for various reasons that we explain below.

## 6.2 Under-sampled multiscale faults

Many field observations suggest that faults feature a complex inner structure (see *Tchalenko and Ambrasey* [1970]; *Klinger et al.* [2005]), consisting of a complex network of sub-faults and secondary brittle structures (like Riedel shears or flower structures, for instance). Some of the substructures may themselves feature a complex inner zone, which thus replicates itself in a more or less self-similar manner. This process necessarily holds down to a lower cutoff scale, which might be of the order of a few rock grain sizes, so that the full fault should ideally be modeled as a closely packed array of a very large number of potentially seismically active subfeatures. This view has been one of the arguments raised by *Ouillon and Sornette* [2011] to justify the use of a Gaussian mixture approach to cluster earthquakes. If we now assume that we can compile a catalog of all events occurring on such a fault, whatever their size and over a very long period of time, with vanishing location uncertainties, then our method would invert correctly the full underlying structure. If the time span of the catalog is much shorter than the typical time scale necessary to activate rupture on every substructure, then the sub-faults will be undersampled by the seismicity process, as most of them would feature very few events, if any. In that case, any method will fail to retrieve the correct structure of the fault zone, and our method would only provide a coarse-grained solution, which may not be necessarily unique. If we now add location uncertainties that are larger than the typical spacing of sub-faults, and sometimes comparable to the spacing of the macro faults, the coarse-graining problem will be transferred to even larger scales, so that the solution to the fit of the full network will not be unique either: different validation techniques will provide different preferred solutions.

In order to illustrate this reasoning, we extended the complexity of the synthetic Landers network of Section 4 down to smaller scales, using an algorithm inspired from the theory of Iterated Function Systems [*Barnsley*, 1988; *Hutchinson*, 1981], a popular technique used to build synthetic fractal sets. In a nutshell, this technique consists in replicating a given fault into another set of randomly rotated and scaled down copies of itself. The set of copies is then used to replace the original fault. The copies are themselves replaced by a similar set of rotated and scaled down copies as well, and so on, down to a given fine-scale resolution. For the sake of simplicity, this segmentation is imposed along the strike of the fault, each sub-plane extending to the same depth as the original fault. An example is shown in Figure 13, and features 220 sub-faults (instead of the 13 original planes). The small-scale structure appears to be very complex, yet the large scale structure is similar to the one presented in Figure 3. We then distribute the same set of 3,103 events over this new set of sub-faults. The network has been generated so that there is, on average, between 10 to 20 events on each segment, but some sub-faults may feature only one or two of them. (Details are given in the electronic supplement).

Using the same method as in Section 4, we generate a new catalog of events providing both their expected locations and their uncertainties.

Focal mechanisms are first chosen as fully compatible with the orientation of the sub-fault to which the event is attributed, before we add a 10° uncertainty on strike, dip and rake. This catalog is then processed by our nonlinear fitting method, using 6,000 runs. This smaller number of runs is a consequence of the much larger duration of individual inversions due to the larger complexity of the dataset, which necessitates a longer time to explore the space of models.

Figure 14 shows the solutions selected by the six validation methods. None of them is able to reconstruct the full set of 220 planes, as expected. All proposed networks feature only 17 to 19 faults, as undersampled sub-faults are indeed merged into simpler structures in order to cluster a sufficient number of events (at least 4, as we imposed). None of the solutions are identical, reflecting the non-uniqueness of the solution provided by the different criteria.

## 6.3 Overfitting, underfitting and validation techniques

The two validation tools based on residuals, i.e. cross validation and BIC, were used in order to avoid problems of overfitting. However, we showed in the previous section that we primarily face a problem of underfitting. This observation necessarily questions the use of such validation strategies for clustering techniques. We also showed that both cross validation and BIC were unable to select the correct solution when a set of background events is superimposed over the more correlated set of earthquakes. This thus leads us to conclude that the use of such criteria is certainly much less adapted to the selection of the correct solution than the use of focal mechanisms, which bring their share of information about the dynamics of the network. Up to now, we only use part of the information contained within focal mechanisms, as we only checked the consistency of the orientation of one of the nodal planes and of the fitting planes. We thus deliberately forgot the rake. In the future, this observation should be included as well in order to better constrain solutions, thus providing a coherent set of slip vectors within the same fault.

## 6.4 Future developments

Our unsupervised clustering technique uses only the spatial information contained within seismicity catalogs. We showed that the model validation criteria derived from focal mechanisms are in better agreement with the true model when dealing with synthetics. A natural idea is then to include more prior seismic information into the clustering procedure itself, like waveform correlation coefficients, focal mechanisms similarities, and so on. However, the design of a cost function able to take account of all those different data necessitates defining a proper weighting strategy. We rather suggest using this extra knowledge to make decisions at decisive steps of the clustering process.

Despite the fact that earthquakes catalogs depict events as point processes, those events indeed define a collection of stress tensors (and their time histories during the rupture process), distributed over a set of finite planar, subplanar or fractal structures. Earthquakes define stress and strain singularities, which obviously interact through stress transmission: earthquakes are triggered by the accumulation of stress at plates' boundaries as well as by stress fluctuations induced by previous events. Earthquakes are also increments of deformation that reveal the development and growth of faults. In return, earthquakes are constrained to occur on such faults. The geometry of the set of events is thus governed both by the applied boundary conditions and the mechanical interactions between events. The overall orientation of faults is mainly governed by the principal directions of the applied boundary stress tensor, while the inner structure and complexity of faults is mainly dominated by interactions between events.

These interactions may propagate over very large distances and time scales, through cascades of domino-effects. Indeed, faults are complex geological structures that are often considered as self-affine surfaces or self-similar aggregates of smaller scale planar features. This means that such objects are significantly correlated over a substantial range of spatial scales. The basic idea we have in mind is that such a correlation must also translate into the dynamical signature of faulting, i.e. the dynamics of the associated earthquakes. Here, we do not use the term 'dynamic' as associated to the temporal distribution of individual events (that is also given in earthquakes catalogs), but to the rupture process of individual events. The idea is that if two events occur within a short spatial distance and belong to the same fault, then there is a 'large' probability that their rupture processes will be similar (which is the basic meaning of correlation). This similarity should, on average, decrease with the distance between events. As all the information we have about the dynamics of faulting is contained within the recorded waveforms, it is thus reasonable to assume that events belonging to the same fault segment will radiate, on average, similar waveforms. Indeed, this similarity is observed and exploited for source model inversion and strong ground motion modeling in using small events as empirical Green's functions (e.g. *Woessner et al.* [2002]).

The most critical and arbitrary step of the clustering algorithm is the one where the locally worst cluster is split into two sub-clusters in order to improve the fit. The chosen cluster is the one with the largest thickness (so that it relies on arguments based on local fit residuals), and the split process is purely random. We suggest that, for a given number of clusters, we may first assess the $\mu_F$ and $\sigma_F$ values for all individual faults. The one(s) with the largest values may then be the chosen ones to be split when increasing the number of planes, so that the splitting is now based on more mechanical grounds. We may also separate those clusters from the rest of the catalog, fit them separately, and put them back into the whole dataset. This would allow fitting separately less complex structures within smaller solution spaces, converging more quickly to a reliable solution. The randomness of the splitting may also be questioned, as we know that the standard k-means algorithm is very sensitive to initial conditions (i.e. the locations of the initial seeds), and that some of them are more optimal than others. In our case, the location, size and orientation of the new planes generated by splitting certainly have a large impact on the reliability of the final solution. Recently, both the k-means++ [*Arthur and Vassilvitskii*, 2007] and the k-means‖ [*Bahmani et al.*, 2012] have been proposed in order to provide better initial conditions to k-means. In k-means++, the first seed is chosen randomly among the data points. All the other seeds are then chosen sequentially from the remaining data points with a probability proportional to their distance squared to the closest previous seed. The k-means‖ is an improvement of k-means++ to deal with large datasets. This technique thus allows one to generate a more or less uniform set of seeds.

The most important obstacle to such clustering techniques is certainly the size of the catalogs to be processed. Up to now, we only considered sets of a few thousands of events, but the full California catalog for instance features up to half a million data points. Processing such large datasets is clearly out of reach of our current algorithm. We may improve it by parallelizing some steps (such as the computations of distances), and also by choosing more efficiently the initial conditions (as outlined above with the k-means++ approach). This limitation to process very large catalogs also holds for other clustering techniques, such as the Gaussian mixture expectation-maximization (EM) approach of *Ouillon and Sornette* [2011]. In the latter paper, a catalog is approximated as a superposition of Gaussian kernels, whose optimal number is determined through a cross-validation strategy. A set of 4,000 events occurring in the Mount Lewis area necessitated about 100 Gaussian kernels for fitting. This large number of objects to fit the data is explained by the fact that the fitting procedure is very sensitive to density fluctuations along a given fault – whereas this is not in the case when fitting with planes. Such a fault, fitted by one single plane following our approach, may require several kernels in the EM approach, which increases the necessary computational resources. We would thus rather use our k-means-based approach to first fit the main faults, then switch to an EM approach to infer more precisely the structure of the fault zones, in the spirit of *Ouillon and Sornette* [2011], who were able to provide a typical segmentation scale – an information of prime importance to model high-frequency ground shaking.

The proposed clustering approach retains the potential to improve the spatial forecasting skills of current forecast models, especially those that attempt short-term near real-time forecasts and are prone to be used for operational earthquake forecasting. Forecast models such as the Short-Term Earthquake Probability (STEP) model [*Gerstenberger et al.*, 2005; *Woessner et al.*, 2010] or the class of epidemic-type earthquake forecast (ETES) models [*Helmstetter et al.*, 2006; *Ogata and Zhuang*, 2006] have been shown to be mostly limited in their spatial predictive skill [*Woessner et al.*, 2011]. Thus, we expect that including the proposed method will improve the forecast skills at least during strong aftershock sequences.

*Acknowledgments*. We would like to thank Egill Hauksson for providing us the Landers catalog and Stefan Wiemer for his valuable comments. This work was funded by the Swiss Science Foundation under Grant 200021-117829 and the European Commission through funds provided within the FP7 project GEISER, grant agreement no. 241321-2.

# References

Arthur, D., and S. Vassilvitskii (2007), k-means plus plus : The Advantages of Careful Seeding, *Proceedings of the Eighteenth Annual Acm-Siam Symposium on Discrete Algorithms*, 1027-1035.

Bahmani, B., B. Moseley, A. Vattani, R. Kumar, and S. Vassilvitskii (2012), Scalable k-means++, *Proc. VLDB Endow.*, *5*(7), 622-633.

Barnsley, M. (1988), Fractals Everywhere, *Academic Press, Inc.*

Bishop, C. M. (2006), *Pattern Recognition and Machine Learning*, Springer.

Bondar, I., S. C. Myers, E. R. Engdahl, and E. A. Bergman (2004), Epicentre accuracy based on seismic network criteria, *Geophys J Int*, *156*(3), 483-496, doi:10.1111/J.1365-246x.2004.02070.X.

Chau, M., R. Cheng, B. Kao, and J. Ng (2006), Uncertain data mining: An example in clustering location data, *Lect Notes Artif Int*, *3918*, 199-204.

Courjault-Radé, P., J. Darrozes, and P. Gaillot (2009), The M = 5.1 1980 Arudy earthquake sequence (western Pyrenees, France): a revisited multi-scale integrated seismologic, geomorphologic and tectonic investigation, *Int J Earth Sci (Geol Rundsch),98*(7),1705-1719, doi:10.1007/s00531-008-0320-5.

Cowie, P. A., D. Sornette, and C. Vanneste (1995), Multifractal Scaling Properties of a Growing Fault Population, *Geophys J Int*, *122*(2), 457-469, doi:10.1111/J.1365-246x.1995.Tb07007.X.

Cowie, P. A., C. Vanneste, and D. Sornette (1993), Statistical Physics Model for the Spatiotemporal Evolution of Faults, *J Geophys Res-Sol Ea*, *98*(B12), 21809-21821, doi:10.1029/93jb02223.

Duda, R. O., P. E. Hart, and D. G. Stork (2001), *Pattern classification*, Wiley.

Felzer, K. R., and E. E. Brodsky (2006), Decay of aftershock density with distance indicates triggering by dynamic stress, *Nature*, *441*(7094), 735-738, doi:10.1038/Nature04799.

Field, E. H., Milner, Kevin R., and the 2007 Working Group on California Earthquake Probabilities (2008), Forecasting California's earthquakes; what can we expect in the next 30 years?, *U.S. Geological Survey*.

Frankel, A. D., D. L. Carver, and R. A. Williams (2002), Nonlinear and linear site response and basin effects in Seattle for the M 6.8 Nisqually, Washington, earthquake, *B Seismol Soc Am*, *92*(6), 2090-2109, doi:10.1785/0120010254.

Gabrielov, A., V. KeilisBorok, and D. D. Jackson (1996), Geometric incompatibility in a fault system, *P Natl Acad Sci USA*, *93*(9), 3838-3842, doi:10.1073/Pnas.93.9.3838.

Gaillot, P., J. Darrozes, P. Courjault-Rade, and D. Amorese (2002), Structural analysis of hypocentral distribution of an earthquake sequence using anisotropic wavelets: Method and application, *J Geophys Res-Sol Ea*, *107*(B10), doi:10.1029/2001jb000212.

Gerstenberger, M. C., S. Wiemer, L. M. Jones, and P. A. Reasenberg (2005), Real-time forecasts of tomorrow's earthquakes in California, *Nature*, *435*(7040), 328-331, doi:10.1038/Nature03622.

Hauksson, E. (2010), Spatial Separation of Large Earthquakes, Aftershocks, and Background Seismicity: Analysis of Interseismic and Coseismic Seismicity Patterns in Southern California, *Pure Appl Geophys*, *167*(8-9), 979-997, doi:10.1007/S00024-010-0083-3.

Hauksson, E., W. Z. Yang, and P. M. Shearer (2012), Waveform Relocated Earthquake Catalog for Southern California (1981 to June 2011), *B Seismol Soc Am*, *102*(5), 2239-2244, doi:10.1785/0120120010.

Helmstetter, A., Y. Y. Kagan, and D. D. Jackson (2006), Comparison of short-term and time-independent earthquake forecast models for southern California, *B Seismol Soc Am*, *96*(1), 90-106, doi:10.1785/0120050067.

Hiemer, S., D.D.Jackson, Q.Wang, Y.Y.Kagan, J. Woesner, J.D.Zechar, and S.Wiemer (2013), A stochastic forecast of California earthquakes based on fault slip and smoothed seismicity, *B Seismol Soc Am*, *103*(2A), doi:10.1785/0120120168.

Husen, S., C. Bachmann, and D. Giardini (2007), Locally triggered seismicity in the central Swiss Alps following the large rainfall event of August 2005, *Geophys J Int*, *171*(3), 1126-1134, doi:10.1111/J.1365-246x.2007.03561.X.

Husen, S., and J.L.Hardebeck (2010), Earthquake location accuracy, *Community Online Resource for Statistical Seismicity Analysis*, doi:10.5078/corssa-55815573.

Husen, S., E. Kissling, N. Deichmann, S. Wiemer, D. Giardini, and M. Baer (2003), Probabilistic earthquake location in complex three-dimensional velocity models: Application to Switzerland, *J Geophys Res-Sol Ea*, *108*(B2), 2077, doi:10.1029/2002jb001778.

Hutchinson, J. E. (1981), Fractals and Self Similarity, *Indiana U Math J*, *30*(5), 713-747, doi:10.1512/Iumj.1981.30.30055.

Klinger, Y., X. W. Xu, P. Tapponnier, J. Van der Woerd, C. Lasserre, and G. King (2005), High-resolution satellite imagery mapping of the surface rupture and slip distribution of the M-W similar to 7.8, 14 November 2001 Kokoxili Earthquake, Kunlun Fault, northern Tibet, China, *B Seismol Soc Am*, *95*(5), 1970-1987, doi:10.1785/0120040233.

Lomax, A., A. Michelini, and A. Curtis (2009), Earthquake Location, Direct, Global-Search Methods, in *Encyclopedia of Complexity and System Science*, edited by R. A. Meyers, pp. 2249-2473, Springer, New York, doi:10.1007/978-0-387-30440-3.

Lomax, A., J. Virieux, P. Volant, and C. Berge-Thierry (2000), Probabilistic earthquake location in 3D and layered models - Introduction of a Metropolis-Gibbs method and comparison with linear locations, *Advances in Seismic Event Location*, 101-134.

Mace, C. G., and K. M. Keranen (2012), Oblique fault systems crossing the Seattle Basin: Geophysical evidence for additional shallow fault systems in the central Puget Lowland, *J Geophys Res-Sol Ea*, *117*, doi:10.1029/2011jb008722.

MacQueen, J. B. (1967), Some Methods for Classification and Analysis of MultiVariate Observations, paper presented at Proc. of the fifth Berkeley Symposium on Mathematical

Statistics and Probability, University of California Press.

Moser, T. J., T. Vaneck, and G. Nolet (1992), Hypocenter Determination in Strongly Heterogeneous Earth Models Using the Shortest-Path Method, *J Geophys Res-Sol Ea*, *97*(B5), 6563-6572, doi:10.1029/91JB03176.

Ogata, Y., and H. C. Zhuang (2006), Space-time ETAS models and an improved extension, *Tectonophysics*, *413*(1-2), 13-23, doi:10.1016/J.Tecto.2005.10.016.

Ouillon, G., C. Castaing, and D. Sornette (1996), Hierarchical geometry of faulting, *J Geophys Res-Sol Ea*, *101*(B3), 5477-5487, doi:10.1029/95jb02242.

Ouillon, G., C. Ducorbier, and D. Sornette (2008), Automatic reconstruction of fault networks from seismicity catalogs: Three-dimensional optimal anisotropic dynamic clustering, *J Geophys Res-Sol Ea*, *113*(B1), doi:10.1029/2007jb005032.

Ouillon, G., and D. Sornette (2011), Segmentation of fault networks determined from spatial clustering of earthquakes, *J Geophys Res-Sol Ea*, *116*, doi:10.1029/2010jb007752.

Ouillon, G., D. Sornette, and C. Castaing (1995), Organisation of joints and faults from 1-cm to 100-km scales revealed by optimized anisotropic wavelet coefficient method and multifractal analysis, *Nonlinear Proc Geoph*, *2*(3-4), 158-177.

Petersen, M. D., T. Q. Cao, K. W. Campbell, and A. D. Frankel (2007), Time-independent and time-dependent seismic hazard assessment for the State of California: Uniform California Earthquake Rupture Forecast Model 1.0, *Seismol Res Lett*, *78*(1), 99-109, doi:10.1785/Gssrl.78.1.99.

Plesch, A., et al. (2007), Community fault model (CFM) for southern California, *B Seismol Soc Am*, *97*(6), 1793-1802, doi:10.1785/0120050211.

Powers, P. M., and T. H. Jordan (2010), Distribution of seismicity across strike-slip faults in California, *J Geophys Res-Sol Ea*, *115*, doi:10.1029/2008jb006234.

Press, W. H., S. A. Teukolsky, W. T. Vetterling, and B. P. Flannery (2007), *Numerical Recipes 3rd Edition: The Art of Scientific Computing*, Third Edition ed., Cambridge University Press.

Rhoades, D. A., and M. W. Stirling (2012), An Earthquake Likelihood Model Based on Proximity to Mapped Faults and Cataloged Earthquakes, *B Seismol Soc Am*, *102*(4), 1593-1599, doi:10.1785/0120110326.

Schwarz, G. (1978), Estimating Dimension of a Model, *Ann Stat*, *6*(2), 461-464, doi:10.1214/Aos/1176344136.

Sornette, D. (1991), Self-Organized Criticality in Plate-Tectonics, *Nato Adv Sci I C-Mat*, *349*, 57-106.

Sornette, D., P. Miltenberger, and C. Vanneste (1994), Statistical Physics of Fault Patterns Self-Organized by Repeated Earthquakes, *Pure Appl Geophys*, *142*(3-4), 491-527, doi:10.1007/Bf00876052.

Sornette, D., and J. Virieux (1992), Linking Short-Timescale Deformation to Long-Timescale Tectonics, *Nature*, *357*(6377), 401-404, doi:10.1038/357401a0.

Tarantola, A., and B. Valette (1982), Inverse Problems = Quest for Information, *Journal of Geophysics*, *50*, 159-170.

Tchalenko, J. S. (1970), Similarities between Shear Zones of Different Magnitudes, *Geol Soc Am Bull*, *81*(6), 1625-&, doi:10.1130/0016-7606(1970)81[1625:Sbszod]2.0.Co;2.

Tchalenko, J.S., and N.N. Ambrasey (1970), Structural Analysis of Dasht-E Bayaz (Iran) Earthquake Fractures, *Geol Soc Am Bull*, *81*(1), 41-&, doi:10.1130/0016-7606(1970)81[41:Saotdb]2.0.Co;2.

Waldhauser, F., and D. P. Schaff (2008), Large-scale relocation of two decades of Northern California seismicity using cross-correlation and double-difference methods, *J Geophys Res-Sol Ea*, *113*(B8), doi:10.1029/2007jb005479.

Werner, M. J., K. Ide, and D. Sornette (2011), Earthquake forecasting based on data assimilation: sequential Monte Carlo methods for renewal point processes, *Nonlinear Proc Geoph*, *18*(1), 49-70, doi:10.5194/Npg-18-49-2011.

Wesson, R. L., W. H. Bakun, and D. M. Perkins (2003), Association of earthquakes and faults in the San Francisco Bay area using Bayesian inference, *B Seismol Soc Am*, *93*(3), 1306-1332, doi:10.1785/0120020085.

Wittlinger, G., G. Herquel, and T. Nakache (1993), Earthquake location in strongly heterogeneous media, *Geophys J Int*, *115*(3), 759-777, doi:10.1111/j.1365-246X.1993.tb01491.x.

Woessner, J., A. Christophersen, J. D. Zechar, and D. Monelli (2010), Building self-consistent, short-term earthquake probability (STEP) models: improved strategies and calibration procedures, *Ann Geophys-Italy*, *53*(3), 141-154, doi:10.4401/Ag-4812.

Woessner, J., S. Hainzl, W. Marzocchi, M. J. Werner, A. M. Lombardi, F. Catalli, B. Enescu, M. Cocco, M.

C. Gerstenberger, and S. Wiemer (2011), A retrospective comparative forecast test on the 1992 Landers sequence, *J Geophys Res-Sol Ea*, *116*, doi:10.1029/2010jb007846.

Woessner, J., D. Schorlemmer, S. Wiemer, and P. M. Mai (2006), Spatial correlation of aftershock locations and on-fault main shock properties, *J Geophys Res-Sol Ea*, *111*(B8), doi:10.1029/2005jb003961.

Woessner, J., M. Treml, and F. Wenzel (2002), Simulation of M-W=6.0 earthquakes in the Upper Rhinegraben using empirical Green functions, *Geophys J Int*, *151*(2), 487-500, doi:10.1046/J.1365-246x.2002.01785.X.

Zechar, J. D., and T. H. Jordan (2008), Testing alarm-based earthquake predictions, *Geophys J Int*, *172*(2), 715-724, doi:10.1111/J.1365-246x.2007.03676.X.

Zechar, J. D., and T. H. Jordan (2010), Simple smoothed seismicity earthquake forecasts for Italy, *Ann Geophys-Italy*, *53*(3), 99-105, doi:10.4401/Ag-4845.

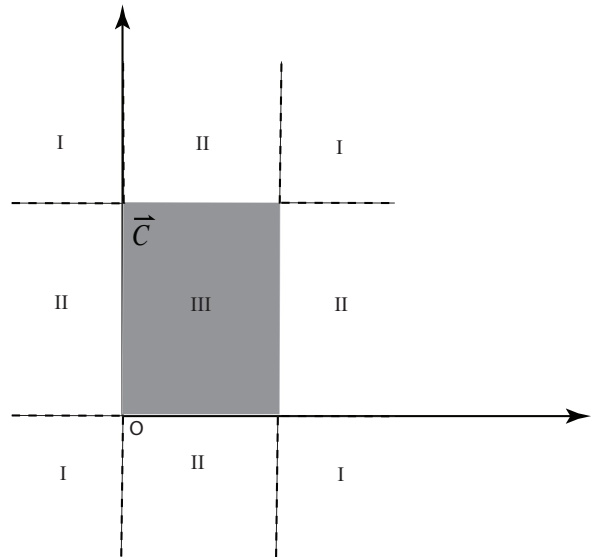

Figure 1: Partition of the 3D space in order to compute the expected square distance of a pdf and a finite plane (shown in grey). The roman indices I, II and III correspond to various approximations of the ESD (see main text, Section 3.3).

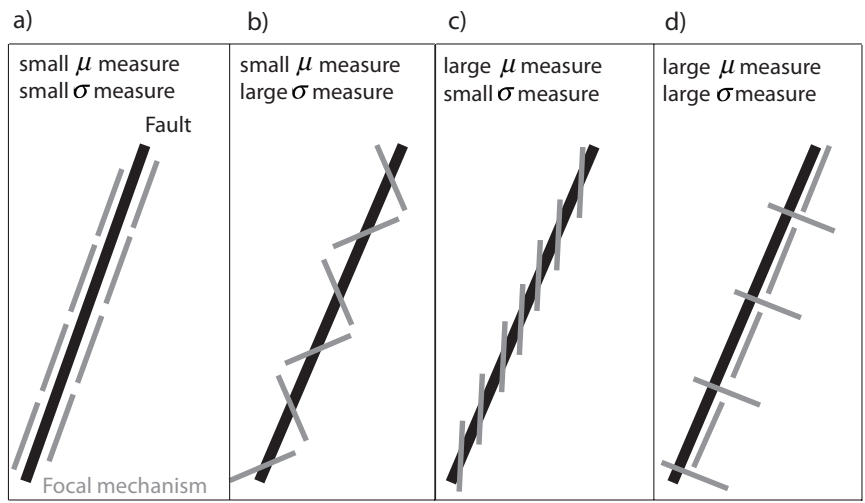

Figure 2: Examples of micro- and macro-structure relationships in fault zones to justify the use of different criteria based on focal mechanisms (see main text, Sections 3.5.2 and 3.5.3). Thick black line denotes general orientation of the fault zone (macro structure); thin gray lines indicate orientation of shorter individual fault planes within the fault zone (micro structure).

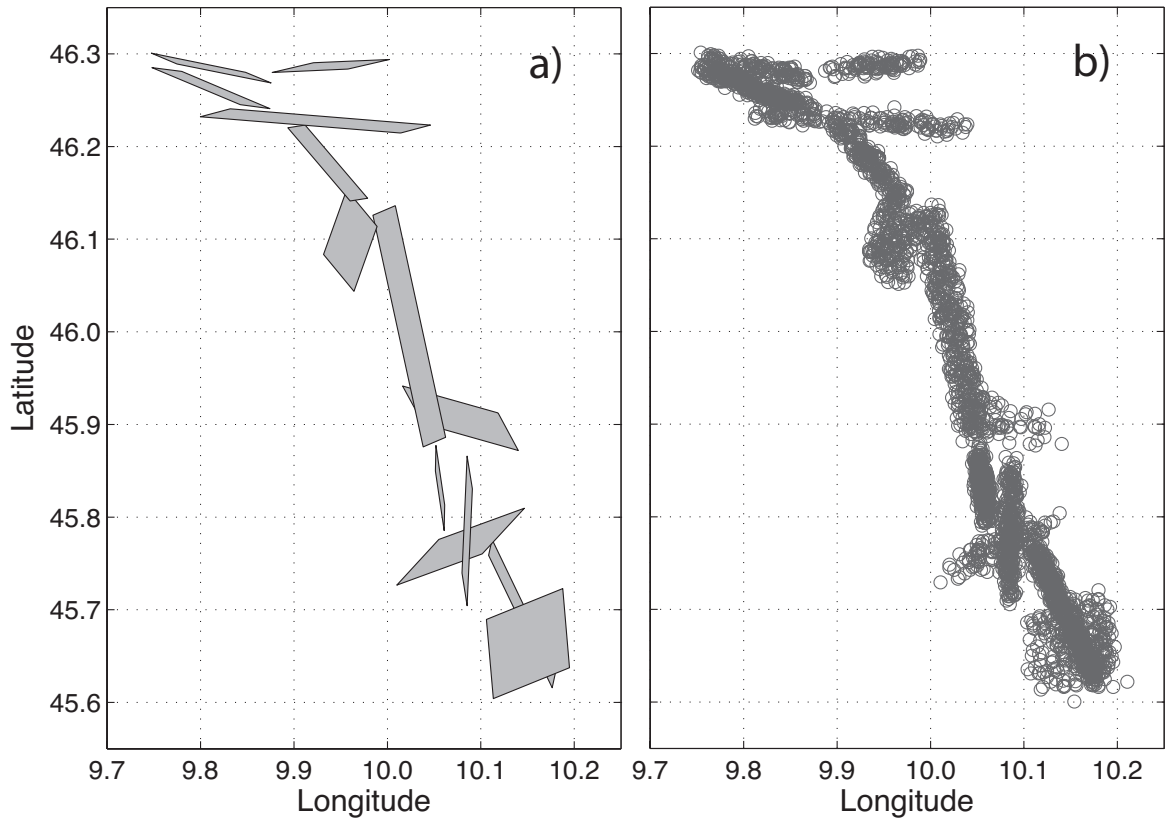

Figure 3: Synthetic data derived from the analysis of [Ouillon et al., 2008] on the Landers fault network. a) Fault network consisting of 13 faults. b) Epicenter map of the synthetic relocated 3,103 events.

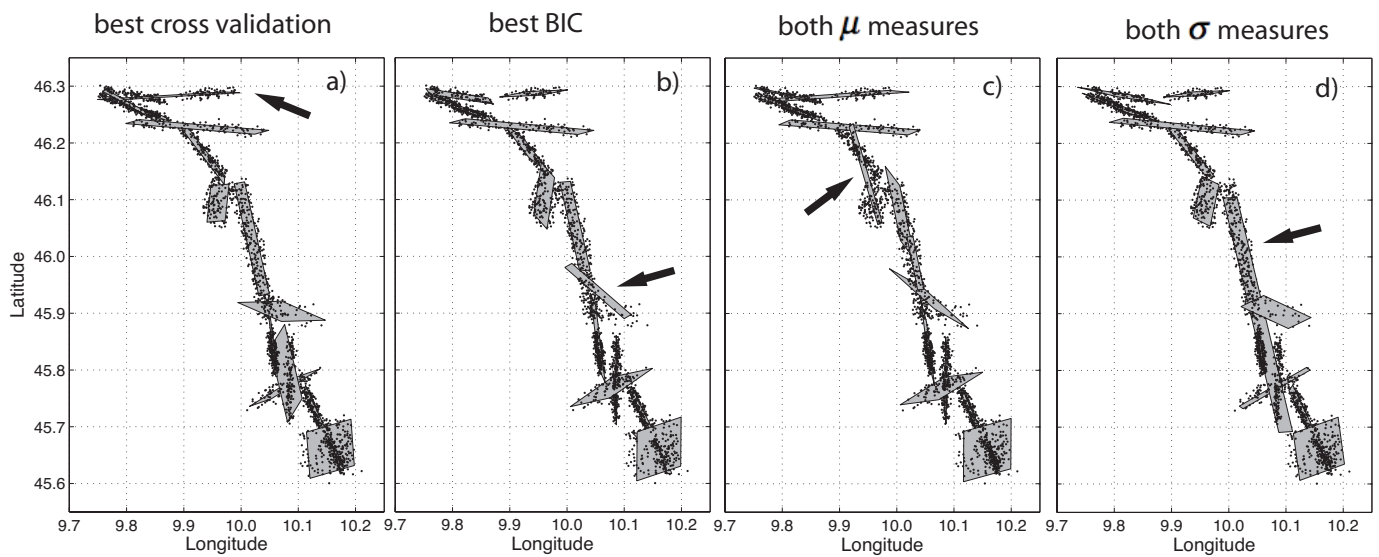

Figure 4: Result of our clustering method applied to the synthetic data consisting of 13 original fault planes and 3,103 events presented in Figure 3. Planes pointed by arrows are spurious faults discussed in Section 4.2.

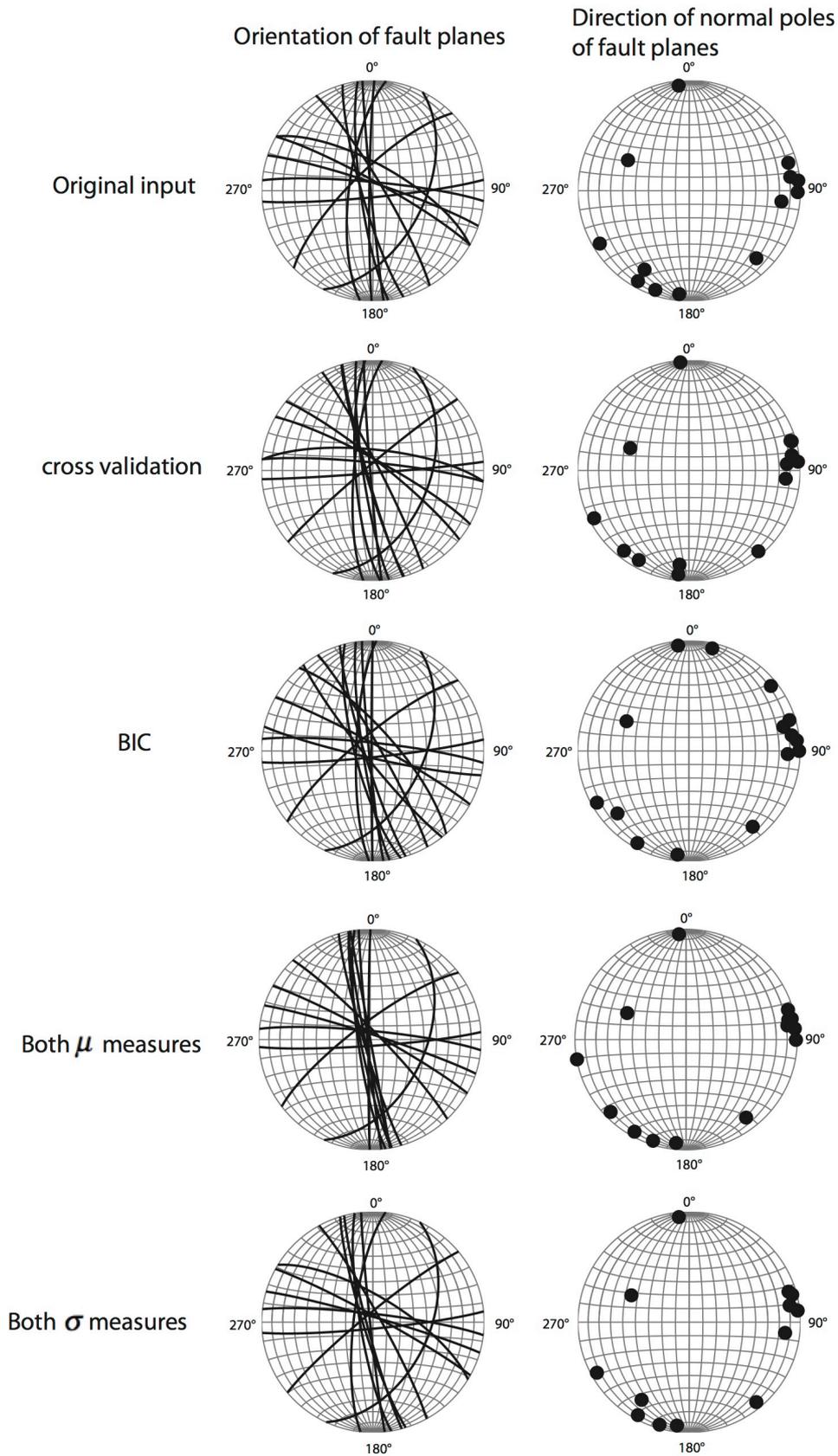

Figure 5: Stereo plots of the original input network and solutions chosen by the six validation criteria. Curves in the left column indicate the orientations of fault traces. Dots in the right column show directions of the normal poles of fault planes.

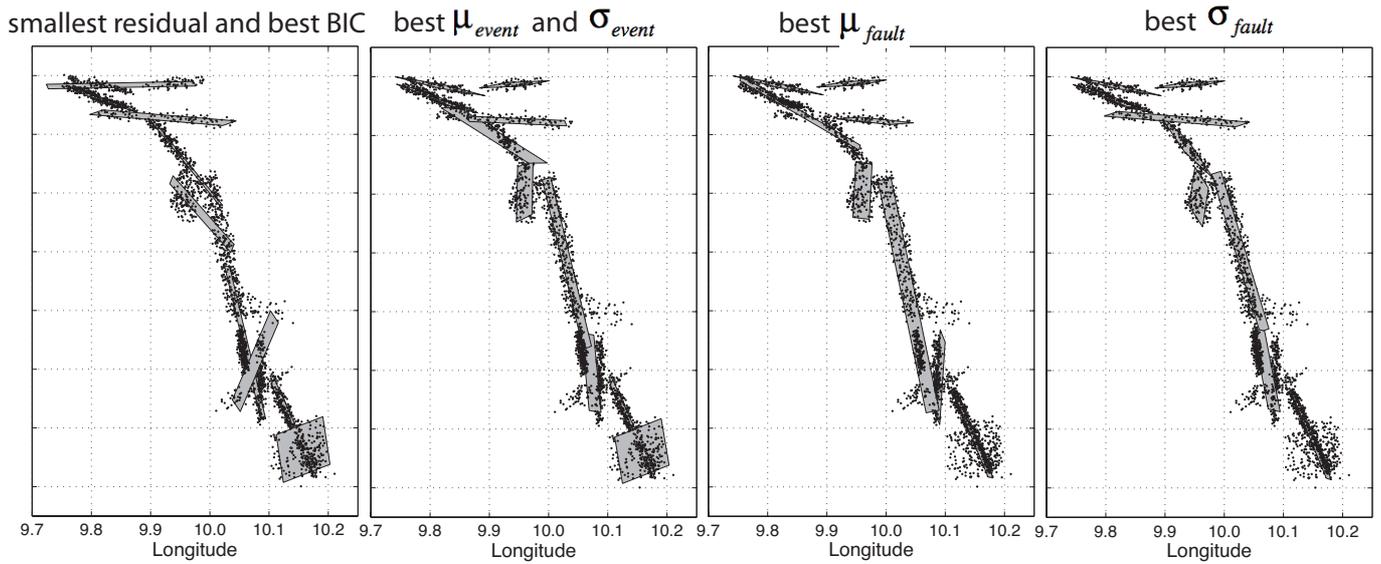

Figure 6: Result of the OADC clustering method of [Ouillon et al., 2008] on a synthetic dataset consisting of 13 original faults and 3,103 events.

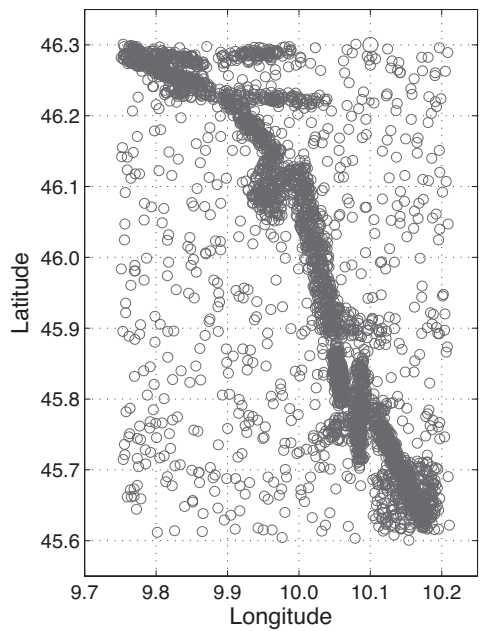

Figure 7: Epicentral map of the synthetic dataset with 20% background events, giving a total of 3,724 events.

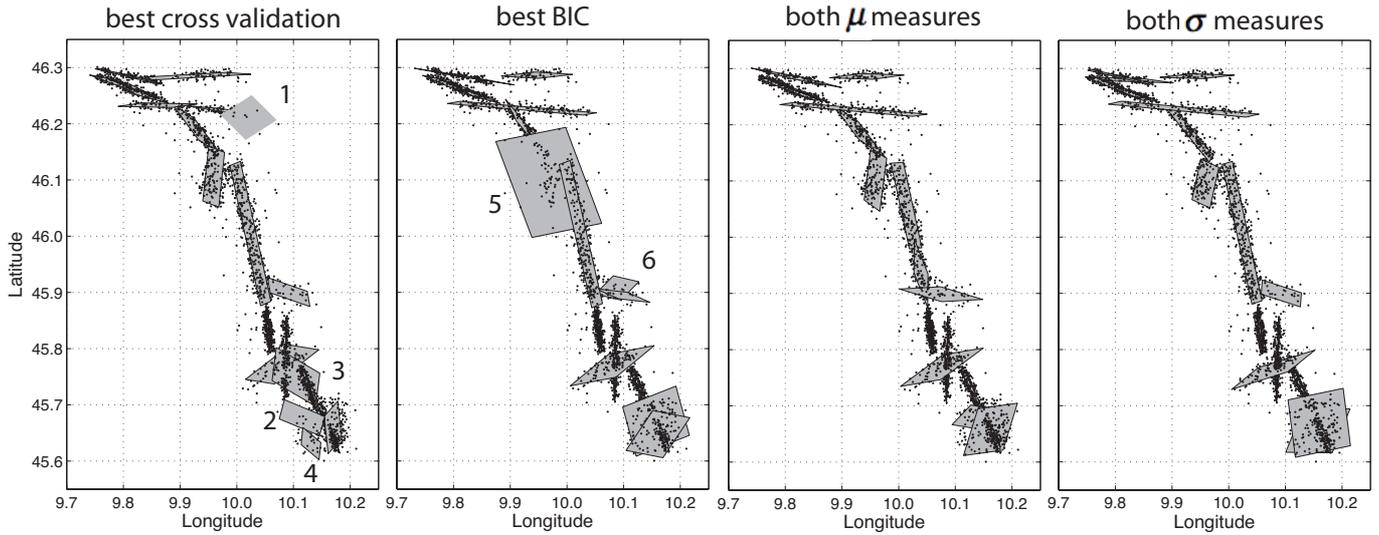

Figure 8: Result of our clustering method applied to the synthetic data set consisting 13 original faults with background seismicity. Solutions chosen by cross validation and BIC feature horizontal planes pointed by numbers are discussed in the text.

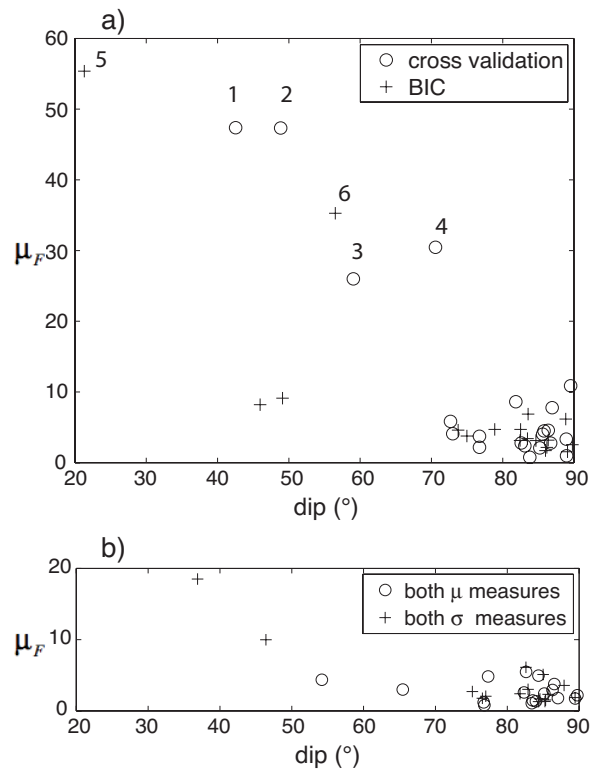

Figure 9: $\mu_F$ value as a function of dip of each reconstructed fault for solutions chosen by different validation criteria. The synthetic data set consists of 13 original faults with background seismicity. Solutions chosen by cross validation and BIC feature horizontal planes with large $\mu_F$ values pointed by numbers and are discussed in the text.

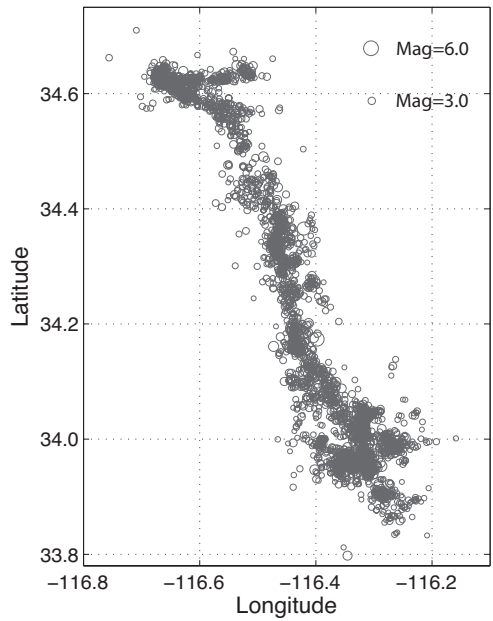

Figure 10: Epicentral seismicity map of the Landers area, 1984-2004. 3360 events were chosen with magnitude > 2, with more than 11 observations, located within an area well-covered by the station network (primary azimuthal gap smaller than 180°, and ratio of the epicentral distance to the closest station over focal depth smaller than 1.5).

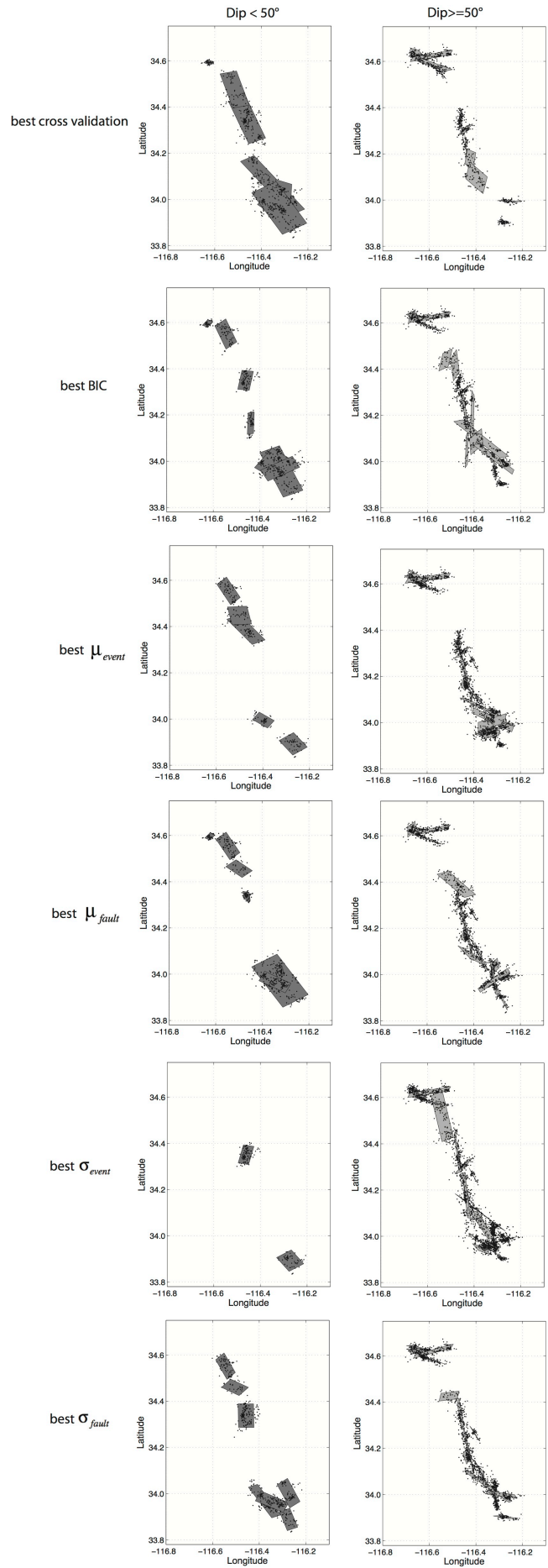

Figure 11: Results of our clustering method applied to the Landers area using the six validation criteria. Results are presented separately for small dipping faults (dip < 50°, left) and large dipping faults (dip >= 50°, right). We assume the solutions using cross-validation and BIC as unrealistic due to the many low-dipping faults in comparison to the tectonically motivated validation measures.

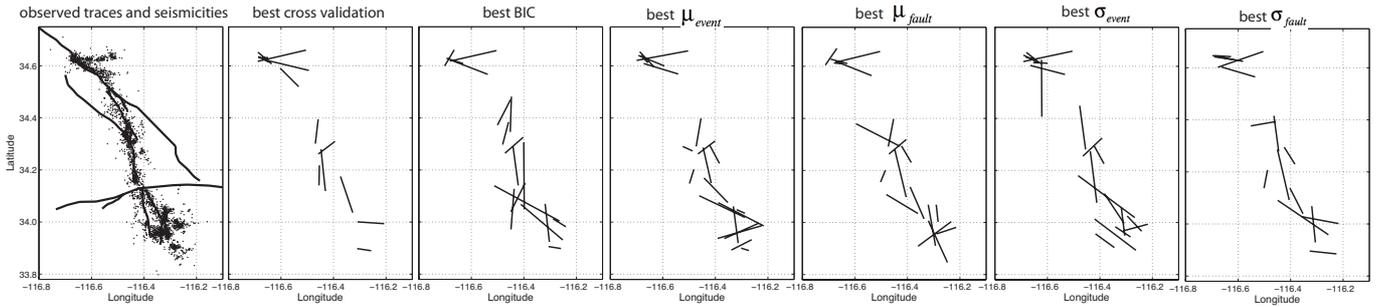

Figure 12: Observed surface traces and seismicity of the Landers area (left plot), and predicted sets of fault traces for each selected reconstructed network.

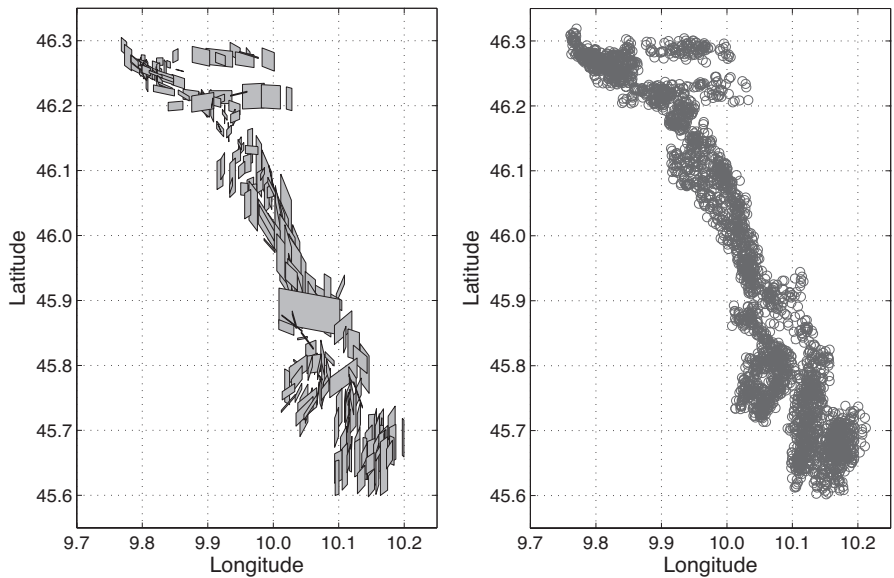

Figure 13: Synthetic multiscale fault network (left) and seismicity (right), consisting respectively of 220 sub-faults and 3,153 events.

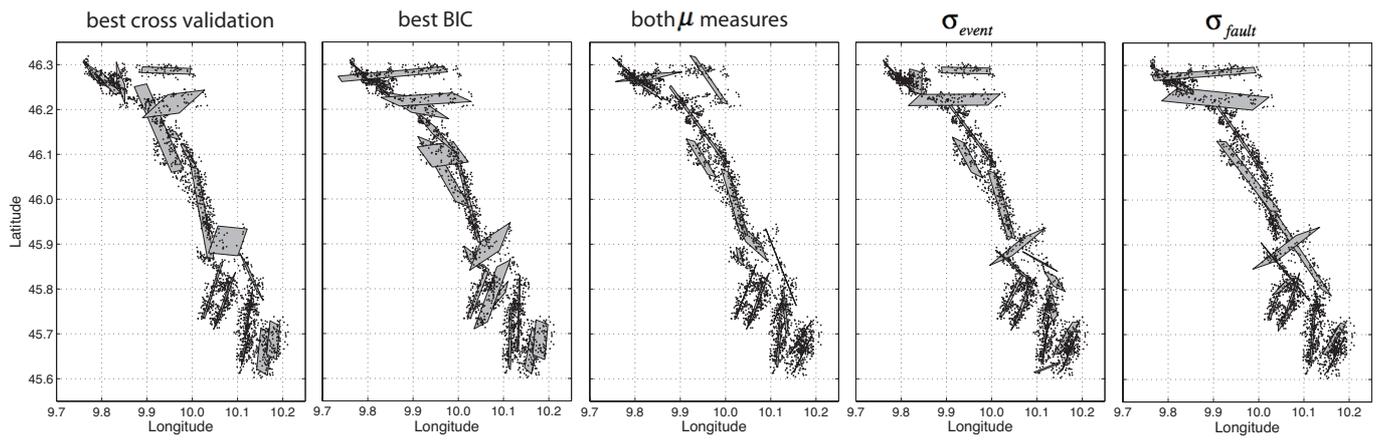

Figure 14: Result of our clustering method applied to the 220 synthetic multiscale fault network consisting of 3103 events. Only 17 to 19 planes were generated.

# Electronic Supplement

## E1 k-means including location uncertainties (uk-means)

The k-means method assumes that the uncertainty of the spatial location of data points is negligible. This assumption holds in disciplines such as image analysis, where the coordinates of the data points are given by red, blue and green color contents at each pixel of a picture. In the case of real physical systems, the story is different. For earthquakes, location uncertainty is an inherent property due to wave arrival time inaccuracy, velocity model errors, station network geometry, or outdated data sources (historical seismicity catalogs, for instance). When taking uncertainty into account, data can no longer be described by a point-process, but by a more or less complex probability distribution function (hereafter pdf).

*Chau et al. [2006]* claim that location uncertainties can significantly affect the results provided by clustering techniques such as k-means. They thus introduce the *uk-means* algorithm (where the 'u' letter stands for 'uncertain'), which incorporates uncertainty information and provides, when considering synthetic samples, more satisfying results than the standard algorithm.

For the general case of a set of objects $\{\overline{O}_1, \overline{O}_2, ..., \overline{O}_n\}$ within an $m$-dimensional space and a set of cluster $\{\overline{C}_1, \overline{C}_2, ..., \overline{C}_k\}$, *k-means* assigns each object to the "closest" cluster barycenter according to the Euclidean distance measure $d(\overline{O}_i, \overline{C}_j) : \mathbb{R}^m \times \mathbb{R}^m \to \mathbb{R}$, where $i = 1,...,n$ and $j = 1,...k$. However, when $\overline{O}_i$ is no longer a point but a pdf, the distance must be estimated differently. *Chau et al. [2006]* propose to use the expected squared distance $ESD(\overline{O},\overline{C})$, defined as the integral of the weighted square norm $\|\overline{O}_i - \overline{C}_j\|^2$ over the whole probability space of $\overline{O}_i$. Denoting the pdf of $\overline{O}_i$ as $f(\cdot)$, we have:

$$\forall \vec{x} \in \overline{O}_i, f(\vec{x}) \geq 0$$
$$\int_{\vec{x} \in \overline{O}_i} f(\vec{x}) d\vec{x} = 1 \quad (1)$$

Then, we define [*Lee et al.*, 2007]:

$$ESD(\overline{O}_i, \overline{C}_j) = \int_{\vec{x} \in \overline{O}_i} \|\vec{x} - \overline{c}_j\|^2 f(\vec{x}) d\vec{x} \quad (2)$$

Where $\overline{c}_j$ is the barycenter of the cluster $\overline{C}_j$. Monte Carlo techniques prove to be too heavy to compute $ESD(\overline{O}_i, \overline{C}_j)$ empirically, especially when dealing with large datasets. A simpler technique consists in using the simple theorem of variance decomposition. *Lee et al. [2007]* thus rewrite Eq. (2) as:

$$ESD(\overline{O}_i, \overline{C}_j) = \int_{\vec{x} \in \overline{O}_i} \|\vec{x} - \overline{k}_i\|^2 f(\vec{x}) d\vec{x} + \|\overline{k}_i - \overline{c}_j\|^2$$
$$= ESD(\overline{O}_i, \overline{k}_i) + \|\overline{k}_i - \overline{c}_j\|^2 \quad (3)$$

where $\overline{k}_i \in \mathbb{R}^m$ is the centroid of the spatial distribution of the uncertain object $\overline{O}_i$ and is defined as $\overline{k}_i = \int_{\vec{x} \in \overline{O}_i} \vec{x} f(\vec{x}) d\vec{x}$. By definition, $ESD(\overline{O}_i, \overline{k}_i)$ is simply the variance of that spatial distribution and can be easily computed once for all from its pdf. The second term in the right hand side in Eq. (3) is simply the square of the distance between two points in a Euclidean space. Using this new distance definition and following the same procedure as standard k-means, data featuring uncertainty information can be processed easily.

Note that when observations come without uncertainties, all pdfs variances are set to 0, so that we recover the classical version of k-means.

# E2 Generating synthetic catalogs with a simple geometry with 3 vertical planes

The general method we propose to generate a synthetic earthquake catalog is the following: we first impose the geometry of the original fault network, which consists in a collection of rectangular planes with variable locations, sizes and orientations. We then assume that all earthquakes occur exactly on those planes and generate P waves. We then compute, assuming a given velocity model, the theoretical travel times between the true hypocenters and a set of stations which locations have been predefined. Random perturbations are added to the waves' arrival times, allowing proceeding to the inverse problem: computing the location of the events as well as their uncertainties. To generate the associated synthetic focal mechanisms, we first assume that the rake of the slip vector on each plane is zero. For each event, the strike and dip are assumed to be identical to the ones of the input plane to which it belongs. We then add a Gaussian random perturbation to the strike, dip and rake of the event with a standard deviation of 10°. Those perturbed angles are then used to compute the strike and dip of the auxiliary plane, thus providing a complete focal mechanism. The inverted location catalog is then fitted with a set of finite planes, using our algorithm for 1,000 clustering runs.

Earthquakes are located using the NonLinLoc software package (*Lomax et al.* [2000], Version 5.2, http://alomax.free.fr/nlloc/). Compared to traditional, linearized approaches, NonLinLoc is superior in that it computes the posterior probability density function (PDF) using nonlinear, global searching techniques. The PDF represents the complete probabilistic solution to the earthquake location problem, including comprehensive information on uncertainty and resolution [*Moser et al.*, 1992; *Tarantola and Valette*, 1982; *Wittlinger et al.*, 1993].

The best solution (which may depend on the validation technique) is then compared to the original input fault network. Note that in real catalogs, the origin of the location uncertainties also lies in the uncertainties about the real velocity model – an ingredient that we neglect here: the sole uncertainties stem from the wave picking process and the geometry of the stations network [*Bondar et al., 2004*].

The first synthetic dataset consists in 4,000 events, uniformly distributed over a network featuring three vertical planes (see Figure 2). Faults A and C have a length (along strike) of 40 km, a width (along dip) of 20 km, and feature 1,000 events each. They share a common strike of 90°E and a common dip of 90°. Fault B has a length of 100 km, a width of 20 km, and features 2,000 events. Its strike is 0°E and its dip is 90°.

We distributed a set of 88 stations on a regular grid with a spatial extent of 240 km by 180 km and a cell size of 20km. For each event, we randomly selected 11 stations out of the complete set of 88 stations as observations, and computed the theoretical arrival times, to which a Gaussian error with a standard deviation of 0.1 s was added to simulate real pickings. A simple 1-D layered velocity model was used. Using NonLinLoc, we generate a synthetic earthquake catalog consisting of 4,000 events characterized by their full pdf.

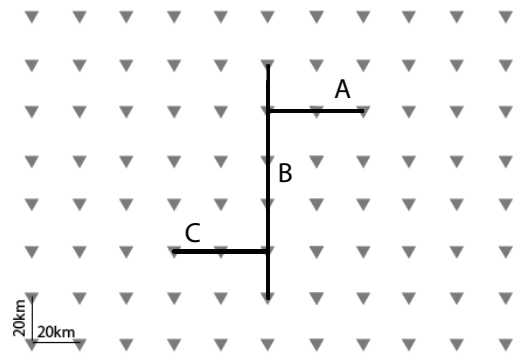

Figure 1: Map view of network design, fault location and earthquake distribution to compute synthetic data. Triangles represent stations (88 in total, 20 km spacing). For each earthquake 11 stations were picked randomly as observations. The lines indicate the fault surface traces along which earthquakes were distributed.

Figure 2a shows the distribution of the 4,000 relocated earthquakes, which are slightly shifted away from the original fault planes. As we use an error-free velocity model and Gaussian picking errors, location uncertainties are mainly controlled by the geometry of the stations network. Events located with a better station network coverage are likely to be characterized by a better location quality. Using the relocated data set, with our clustering technique we generated 1,000 reconstruction solutions.

The solutions are chosen by cross validation, using both $\mu$ and $\sigma_{event}$ measures, and consist of three planes with similar structure, which are shown in Figure 2b. Table 1 lists the parameters of the true and the reconstructed faults. All those solutions show a nice agreement with the true fault network. Figure 2c shows the solution chosen by BIC. Fault B is divided into two sub-faults at the intersection with fault C. From other tests we performed, we also noticed that, when faults cross each other, our model has difficulties in deciding which plane one event belongs to. Yet, the structure is still nicely inverted. Figure 2d displays the solution chosen by $\sigma_{fault}$. One small fault is generated at the northern edge of fault B. This comes from the fact that locations quality close to the northern and southern edges is not as good as in other parts due to a poorer station coverage. We also performed similar tests on catalogs generated with different numbers of observations and different Gaussian picking errors, and obtained similar results.

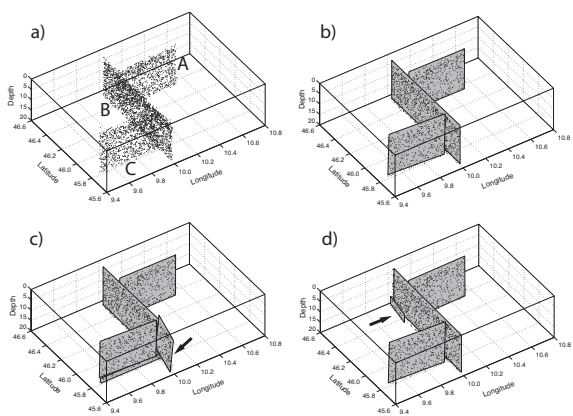

Figure 2: a) Distribution of 4,000 relocated hypocenters located on three vertical faults A, B and C. b) Reconstructed structures from cross validation, using both $\mu$ and $\sigma_{event}$ measures. Three vertical faults are clustered. c) Result from BIC: Fault B is divided into two faults in the southern part. d) Result using $\sigma_{fault}$. One small fault is generated at the northern edge of fault B.

# E3 Generating multiscale synthetic fault catalogs

The method we use to generate multiscale fault networks is largely borrowed from the concept of Iterated Function Systems (hereafter IFS), proposed by *Hutchinson* [1981] and popularized by *Barnsley* [1988], which provide a basic algorithm to generate deterministic fractal objects. IFS consist in replacing a given large scale Euclidean object by a series of replications of itself at smaller and smaller scales. In our case, this consists in segmenting the fault at smaller and smaller resolutions.

We shall first consider a vertical fault, over which events are distributed. The case of a fault with arbitrary strike and dip is solved by simply performing the necessary rotations of such a vertical fault. The fault is chosen such that its length is 1. The general case is solved by a simple scaling up (or down) to the real length of the fault.

We first consider only the trace of the fault, which is its intersection with the free surface, and will perform the segmentation along its strike. The x axis is chosen to stand along the fault's strike, the y axis is normal to the fault, and the z axis is pointing downwards.

**STEP 1**

Consider a fault $L_1$ with length 1 which extremities are [0,0] and [1,0].

**STEP 2**

Define a group of linear applications, for example 10 different functions $F_i$, with i=1,2,...10. $F_i$ is a linear transformation function which reads:

x' = $A_i$ x – $B_i$ y + $C_i$

y' = $B_i$ x + $A_i$ y + $D_i$

with $A_i^2 + B_i^2 = Q_i^2$.

The coefficients A, B, C and D are chosen such that this application transform a given fault segment into a downscaled, slightly rotated and offset copy of itself (so that Q < 1). The various parameters of the set of functions may be chosen by hand or randomly.

Table 1: Parameters of the true and the reconstructed fault networks discussed in the text.

| | Fault | Center of Planes | | | Orientation | | Dimension | |
|---|---|---|---|---|---|---|---|---|
| | | Long | Lat (°) | Depth | Strike (°) | Dip (°) | Length (°) | Width (°) |
| Input | A | 10.27 | 46.36 | 10.0 | 90.00 | 90.00 | 40.0 | 20.0 |
| | B | 10.00 | 46.09 | 10.0 | 0.00 | 90.00 | 100.0 | 20.0 |
| | C | 9.74 | 45.82 | 10.0 | 90.00 | 90.00 | 40.0 | 20.0 |
| Cross validation | A' | 10.27 | 46.36 | 10.4 | 269.88 | 89.98 | 37.7 | 19.3 |
| | B' | 10.00 | 46.09 | 10.2 | 0.05 | 89.98 | 100.7 | 18.3 |
| | C' | 9.74 | 45.82 | 10.1 | 269.90 | 89.99 | 38.5 | 19.2 |
| BIC | A' | 10.27 | 46.36 | 10.37 | 89.96 | 89.98 | 37.9 | 19.3 |
| | $B_1'$ | 10.00 | 46.16 | 10.2 | 180.03 | 89.98 | 82.1 | 18.3 |
| | $B_2'$ | 10.00 | 45.73 | 9.87 | 359.04 | 89.85 | 21.4 | 17.5 |
| | $C_1'$ | 9.77 | 45.82 | 19.22 | 90.33 | 48.01 | 42.0 | 2.2 |
| | $C_2'$ | 9.74 | 45.82 | 9.4 | 89.9 | 90.0 | 38.7 | 17.3 |
| Both $\mu$ | A' | 10.28 | 46.36 | 10.3 | 89.83 | 89.96 | 36.9 | 19.2 |
| | B' | 10.00 | 46.09 | 10.2 | 0.10 | 89.93 | 101.1 | 18.3 |
| | C' | 9.74 | 45.82 | 10.1 | 269.90 | 89.99 | 38.5 | 19.2 |
| $\sigma_{event}$ | A' | 10.27 | 46.36 | 10.3 | 269.87 | 89.96 | 37.4 | 19.2 |
| | B' | 10.00 | 46.09 | 10.2 | 0.04 | 89.96 | 100.7 | 18.3 |
| | C' | 9.74 | 45.82 | 10.2 | 89.90 | 89.97 | 38.6 | 19.2 |
| $\sigma_{fault}$ | A' | 10.27 | 46.36 | 10.4 | 269.86 | 89.98 | 38.1 | 19.3 |
| | $B_1'$ | 10.00 | 46.47 | 18.3 | 359.73 | 86.96 | 20.1 | 4.1 |
| | $B_2'$ | 10.00 | 46.08 | 10.0 | 180.06 | 89.97 | 99.7 | 17.9 |
| | C' | 9.73 | 45.82 | 10.2 | 89.91 | 90.00 | 38.4 | 19.2 |

**STEP 3**

(i) Choose randomly one of the N functions $F_i$ previously defined and apply it to $L_1$ so that one gets a new segment $S_1$ and its extremities

(ii) Repeat step (i) a few times (p times, for instance, with p small). Doing so, a set of new small segments $S_j$, with j=1,…,p, is generated. Store the coordinates of their extremities. Remove the original, large scale segment $L_i$.

(iii) The new dataset now consists in the set {$S_j$}. Apply steps (i-ii) to each of its members.

(iv) Iterate step (iii) a few times so that, at each iteration, the full set of newly created segments {$S_j$} replaces the previous one. The total number of segments thus increases after each iteration while their sizes decrease.

**STEP 4**

Rescale the final lengths of the segments so that the extent of the set fits within [0;1] along its average direction.

**STEP 5**

Apply steps 1 to 4 to generate a different segmented fault for each fault of the catalog. Rotate the segmented fault accordingly so that its average strike and dip fit with the original one.

The previous algorithm thus provides a segmentation of the original fault along its strike, but not along its dip where we leave its structure intact. But a similar process can be implemented along that direction too. Alternatively, for the sake of simplicity, we can achieve a 3D structure by extending each subplane to the same depth as the original fault. Their knowledge allows one to locate some events on those segments and build their focal mechanisms. If the total number of events generated on the whole fault is very large, then each segment will feature enough events to be fully identifiable from them. If the number of events is too small, each segment will be undersampled by the synthetic seismicity catalog, resulting in a noisy multiscale subnetwork.

Figure 3 shows an example to generate synthetic multiscale synthetic faults following the approach we discussed above. The original fault shown in Figure 3a is a structure with strike=172°, dip=81° and dimensions 28km×3km. There are 446 events located on it. In order to generate the set of multiscale synthetic faults, we randomly generated at each iteration 2 linear transformation functions to build smaller scale segments. We constrain the linear transformations so that the distribution of small faults is still along the strike direction. After 5 iterations, we finally generated a multiscale set of 32 fault segments. The number of events located on each small fault depends on its size and ranges from 4 to 71.

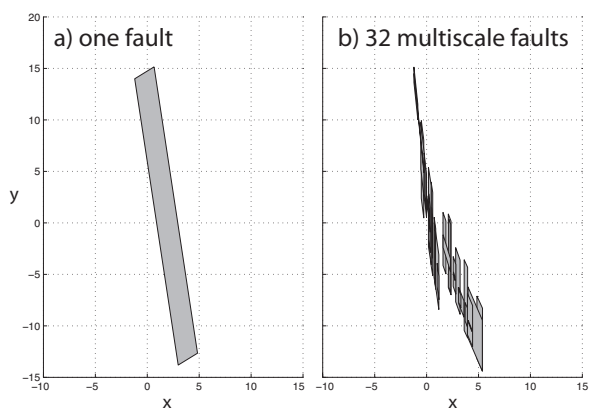

Figure 3: The original fault shown on the left has a strike=172°, dip=81° structure with dimension 28km×3km. Based on it, the multiscale faults structure consisting of 32 small segments (right) was generated following the approach discussed in the text. There are 446 events in total located on the original fault (left). For the multiscale faults, the number of events located on each fault ranges from 4 to 71 events.

# References

Barnsley, M. (1988), Fractals Everywhere, *Academic Press, Inc.*

Bondar, I., S. C. Myers, E. R. Engdahl, and E. A. Bergman (2004), Epicentre accuracy based on seismic network criteria, *Geophys J Int*, *156*(3), 483-496, doi:10.1111/J.1365-246x.2004.02070.X.

Chau, M., R. Cheng, B. Kao, and J. Ng (2006), Uncertain data mining: An example in clustering location data, *Lect Notes Artif Int*, *3918*, 199-204.

Hutchinson, J. E. (1981), Fractals and Self Similarity, *Indiana U Math J*, *30*(5), 713-747, doi:10.1512/Iumj.1981.30.30055.

Lee, S. D., B. Kao, and R. Cheng (2007), Reducing UK-Means to K-Means, in *Proceedings of the Seventh IEEE International Conference on Data Mining Workshops*, edited, pp. 483-488, IEEE Computer Society, doi:10.1109/icdmw.2007.81.

Lomax, A., J. Virieux, P. Volant, and C. Berge-Thierry (2000), Probabilistic earthquake location in 3D and layered models - Introduction of a Metropolis-Gibbs method and comparison with linear locations, *Advances in Seismic Event Location*, 101-134.

Moser, T. J., T. Vaneck, and G. Nolet (1992), Hypocenter Determination in Strongly Heterogeneous Earth Models Using the Shortest-Path Method, *J Geophys Res-Sol Ea*, *97*(B5), 6563-6572, doi:10.1029/91JB03176.

Tarantola, A., and B. Valette (1982), Inverse Problems = Quest for Information, *Journal of Geophysics*, *50*, 159-170.

Wittlinger, G., G. Herquel, and T. Nakache (1993), Earthquake location in strongly heterogeneous media, *Geophys J Int*, *115*(3), 759-777, doi:10.1111/j.1365-246X.1993.tb01491.x.